\documentclass{ar-1col}

\usepackage{url}
\usepackage{natbib}
\usepackage{longtable}
\usepackage{graphics, graphicx}
\usepackage{subcaption}

\jname{Xxxx. Xxx. Xxx. Xxx.}
\jvol{AA}
\jyear{YYYY}
\doi{10.1146/((please add article doi))}

\newcommand{\lsim}{{\;\raise0.3ex\hbox{$<$\kern-0.75em\raise-1.1ex\hbox{$\sim$}}\;}}
\newcommand{\gsim}{{\;\raise0.3ex\hbox{$>$\kern-0.75em\raise-1.1ex\hbox{$\sim$}}\;}}

\begin{document}

\markboth{Gezari}{Tidal Disruption Events}

\title{Tidal Disruption Events}

\author{Suvi Gezari,$^{1,2}$ 
\affil{$^1$Department of Astronomy, University of Maryland, College Park, MD, 20742-2421; email: suvi@astro.umd.edu} \affil{$^2$Joint Space-Science Institute, University of Maryland, College Park, MD, 20742-2421}}

\begin{abstract}
The concept of stars being tidally ripped apart and consumed by\\
a massive black hole (MBH) lurking in the center of a galaxy first \\
captivated theorists in the late 1970's.  The observational evidence \\
for these rare but illuminating phenomena for probing otherwise\\ 
dormant MBHs, first emerged in archival searches of the soft X-ray\\
ROSAT All-Sky Survey in the 1990's; but has recently
accelerated \\
with the increasing survey power in the optical time domain, with\\
tidal disruption events (TDEs) now regarded as a class of optical\\
nuclear transients with distinct spectroscopic features.  \\
Multiwavelength observations of TDEs have revealed \\
panchromatic emission, probing a wide range of scales, from the \\
innermost regions of the accretion flow, to the surrounding\\ 
circumnuclear medium.  I review the
current census of 56 TDEs \\
reported in the literature, and their observed properties can be\\ summarized as follows:

\noindent $\bullet$ The optical light curves follow a power-law decline from peak that\\ 
scales with the inferred central black hole mass as expected for the\\ 
fallback rate of the stellar debris, but the rise time does not.  \\
$\bullet$ The UV/optical and soft X-ray thermal emission come from\\
different spatial scales, and their intensity ratio has a large \\
dynamic range, and is highly variable, providing important clues\\ as to what is
powering
 the two components.\\
$\bullet$ They can be grouped into three spectral classes, and those with \\
Bowen fluorescence line emission show a preference for a hotter and\\
more compact line-emitting region, while those with only He II \\
emission lines are the rarest class.\\

\end{abstract}

\maketitle

\tableofcontents

\section{INTRODUCTION}

Tidal disruption events (TDEs) originated in the late 1970s as a {\it theoretical} concept; a dynamical consequence of massive black holes (MBHs) speculated to be at the centers of most galaxies \citep{hills1975, lidskii1979}.  A star's orbit could bring it close enough to the MBH to be disrupted or captured by the black hole, depending on the relative size of the tidal disruption radius to the black hole event horizon.  When a star is disrupted outside the event horizon, which for a solar type star occurs for $M_{\rm BH} \lsim 10^{8} M_\odot$, a luminous flare of radiation is expected from the bound fraction of the tidal debris that falls back onto the black hole, and circularizes to form an accretion disk \citep{Rees1988, Phinney1989, Evans1989, Ulmer1999}.  This electromagnetic signal was suggested at the time to be one of the best probes for dormant MBHs lurking in the galaxy centers, when direct dynamical measurements of black hole mass are not feasible.   

TDEs also act as a cosmic laboratory to study the real-time formation of an accretion disk and jet.  In a TDE, a previously dormant black hole goes through a dramatic transformation, with a sudden new influx of gas, and potentially magnetic fiux, available from the disrupted star for accretion.  Given that the timescale for fallback and circularization of the stellar debris streams is as short as months (see \S \ref{sec:time}), TDEs allow us to witness the process of the assembly of the nascent accretion disk, and and the launching of any accompanying outflow or jet.  
Once the accretion process begins, we can also probe the gas density of the inner regions of galaxy nuclei from observing the photoionization and interaction of jets and outflows with its circumnuclear environment.

The rate at which stars are destined to be tidally disrupted is well determined by the mass of the central black hole, and the nuclear stellar density and orbital distribution.  Analytical and numerical calculations for stars scattering into the MBH ``loss cone'' \citep{Frank1976}, the angular momentum phase-space for which $R_{\rm p} < R_{\rm T}$, where $R_{\rm p}$ is the pericenter of the star's orbit, are in general agreement for a rate that ranges from $10^{-4}-10^{-5}$ yr$^{-1}$ galaxy$^{-1}$ \citep{Magorrian1999, Wang2004, Stone2016b, Brockamp2011}, with rates as high as $10^{-3}$ yr$^{-1}$ galaxy$^{-1}$ in galaxies with steep nuclear density profiles such as nucleated dwarf galaxies \citep{Wang2004} and post-starburst (``E+A'') galaxies \citep{Stone2016}.  The spin distribution of MBHs can also leave an imprint on the TDE rate, especially at masses close to   the maximum mass for which the tidal disruption radius is still outside the black hole's event horizon, $M_{\rm max}$, which increases from  $\approx 1 \times 10^{8} M_{\odot}$ for a nonspinning ("Schwarzschild") BH up to $7 \times 10^{8} M_\odot$  for a maximally spinning ("Kerr") black hole, due to the effects of the shrinking event horizon with increasing spin \citep{Beloborodov1992, Kesden2012}.  

\begin{figure}[t]
\includegraphics[width=3in,,trim=1cm 0 4cm 14cm, clip]{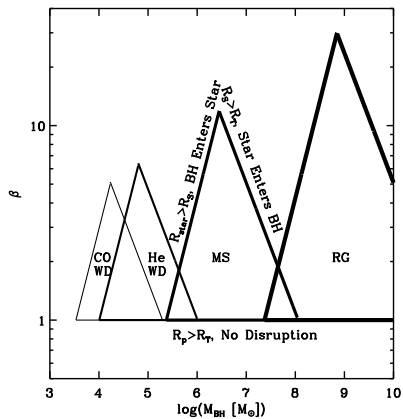}
\caption{\footnotesize Allowable region for the tidal disruption of stars representative  of different evolutionary states, a $0.6 M_\odot$ carbon oxygen white dwarf (CO WD), a $0.17 M_\odot$ helium white dwarf (He WD), a $1 M_\odot$ main-sequence star (MS), and a $1.4 M_\odot$ red giant (RG), bounded by the conditions that $R_p < R_T, R_\star < R_S$, and $R_T > R_S$ for a TDE to be observable, as a function of black hole mass ($M_{\rm BH}$) and $\beta$ is the strength of the tidal encounter ($\beta \equiv R_{\rm T}/R_{\rm p}$).  Diagram inspired by \cite{Rosswog2009}.}
\label{fig:triangle}
\end{figure}

The presence of a binary MBH, as a product of a recent galaxy merger, can also temporarily enhance (suppress) the TDE rate due to the pumping up (ejection) of stars on radial orbits, until the binary MBHs inspiral, and the loss cone is refilled by two-body interactions \citep{Milosavljevic2003, Lezhnin2016}.  Finally, the mere presence of a central black hole can be tested by comparing to the TDE rates expected for galaxy mass functions with different galaxy-mass dependent black hole occupations fractions \citep{Stone2016b}, which may reveal the formation mechanism of the MBHs themselves \citep{Greene2012}.  

Given these sensitivities of the TDE rate to the mass, spin, binarity and occupation fraction of MBHs, they serve as excellent probes of MBH demographics.  
Finally, it should be noted that the demographics, internal structure, and evolutionary state of the unlucky stars themselves can be revealed from the timescales and chemical composition of the observed TDEs \citep{Syer1999, Lodato2009, MacLeod2012, Guillochon2013, Kochanek2016, Ryu2020, Law-Smith2020}.  In Figure \ref{fig:triangle}, I plot the distinct regions of the parameter space of black hole mass and impact parameter that TDEs can probe, from the disruption of white dwarfs by intermediate-mass black holes to the disruption of red giants by the most massive black holes known. 

Despite the strong theoretical motivations for using TDEs to probe MBHs, and the current census of 56 TDEs reviewed here, progress for using TDEs as probes of black hole demographics has been severely hampered by our lack of understanding of the emission mechanism powering TDEs.  Indeed, in the last decade, observations of TDEs have outpaced theoretical modeling, with many important aspects of TDE properties still yet to be definitively and consistently accounted for with a physically motivated model for the tidal disruption debris and its eventual assembly into an accretion disk. However, the goal of this review is to present our latest understanding of the {\it observed} properties of TDEs, with the hopes that a theoretical understanding will soon follow after a global view of the TDE population.

\section{OBSERVED CANDIDATES}
\begin{figure}[t]
\begin{subfigure}[t]{3in}
\centering
\includegraphics[width=3in,trim=1cm 0 4cm 14cm, clip]{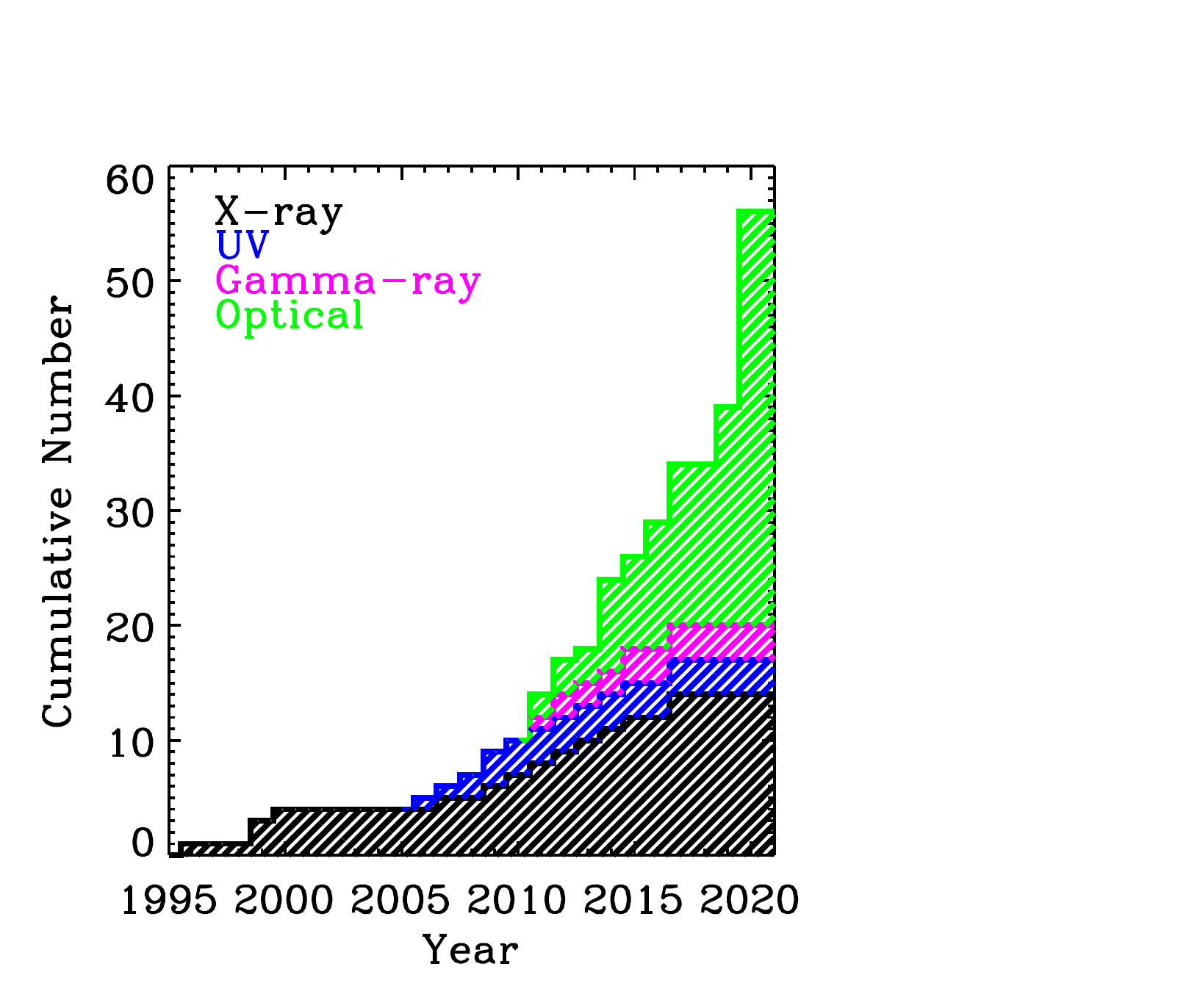}
\end{subfigure}
\begin{subfigure}[t]{3in}
\centering
\includegraphics[width=3in,trim=1cm 0 4cm 14cm, clip]{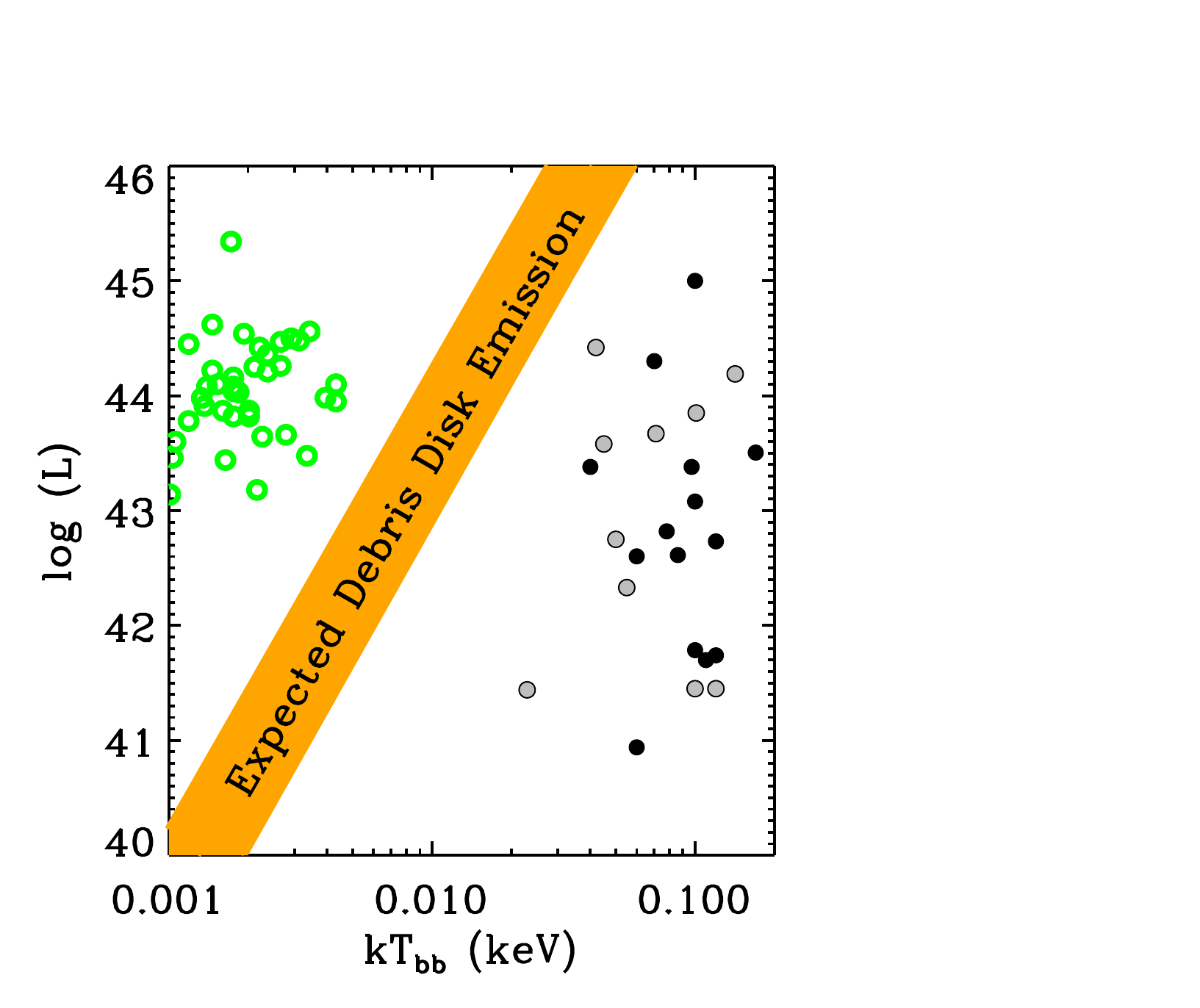}
\end{subfigure}
\caption{\footnotesize {\it Left}: Cumulative histogram of TDEs reported in the literature, color-coded by the wavelength in which they were discovered, X-ray (black), UV (blue), gamma-ray (purple), and optical (green).  {\it Right}: Peak luminosity versus blackbody temperature for 56 TDEs reported in the literature, color coded by the wavelength in which they were discovered, UV/optical (green), X-ray (black), including the 10 of the UV/optically selected TDEs with detected X-ray components (grey).  The UV/optical luminosities are calculated for the entire blackbody, while the X-ray luminosities are only for the 0.3-2 keV band, but should account for most of the bolometric luminosity given the extremely soft temperatures observed.  The region of expected thermal emission from a circularized debris disk formed from the tidal disruption of a solar-type star by a $10^6-10^8 M_\odot$ black hole is shown in orange.  Note that neither of the two components are in agreement with emission expected from a simple debris disk.}
\label{fig:tde_year}
\end{figure}

We are still at a point in the field of TDEs, that the well studied candidates can be listed individually.  In Table \ref{table1} I list the 56 TDE candidates that have been reported in the literature, including the name of the survey that discovered the transient, the waveband of the survey (ultraviolet, optical, X-ray, or gamma-ray), the peak luminosity and the characteristic thermal blackbody temperature at peak, and the reference associated with those measurements.   
All of these nuclear transients have been shown to not harbor an active galactic nucleus (AGN) or a supernova (SN) in the host galaxy nucleus, with a TDE as the most viable explanation for the flaring source.  In principle TDEs should occur around massive black holes that have an AGN accretion disk, and at least one strong candidate has been reported in a Narrow-Line Seyfert 1 galaxy \citep{Blanchard2017}.  TDEs in an AGN host may even produce interesting signatures from the interaction with the debris streams with the pre-existing AGN disk that could distinguish them from TDEs around quiescent black holes \citep{Chan2019}.  However, there are too many ways in which observations of a variable AGN could be mistook with the signatures of a TDE flare, and so I avoid these candidates to ensure the purity of the TDE census.
In Figure \ref{fig:tde_year} we show the cumulative number of TDEs reported in the literature as a function of time and color-coded by the survey waveband.  Once can see how wide-field optical surveys have played an important role in the last decade in discovering TDEs, and account for almost 2/3 of all the reported TDEs to date.

One of the most distinguishing characteristics of TDEs is that their continuum is very well described by a thermal blackbody.  However, as can be seen in Figure \ref{fig:tde_year}, the temperature distribution of TDEs is bimodal, and seemingly dependent on the waveband in which they were discovered.  However, there are now 10 UV/optically selected TDEs for which a soft X-ray component was also detected in follow-up X-ray observations (see \S \ref{sec:xray}).  We plot their blackbody temperatures in gray, and see that their characteristic temperatures fall within the X-ray selected TDE distribution, indicating that these two components are in fact physically distinct.  In \S \ref{sec:opt_xray} we discuss how the UV/optical and X-ray thermal components are related, and present an important population of nearby TDEs for which {\it both} components have been detected, and reveal clues about their physical nature from their relative temporal evolution.

\subsection{Soft X-ray Candidates}


The first TDE candidates emerged from archival searches of the ROSAT All Sky Survey (RASS) conducted in $1990-1991$, whose coverage of 20\% of the sky in the 0.1-2.4 keV band, when compared to follow-up pointed observations the with Position Sensitive Proportional Counter (PSPC), revealed luminous ($10^{41-44}$ erg s$^{-1}$), extremely soft ($\Gamma > 3$) X-ray outbursts from four galaxies with no previous evidence of an AGN, that implied a rate of $\sim 1 \times 10^{-5}$ yr$^{-1}$ per galaxy, consistent with theoretical dynamical predictions for the rates of TDEs \citep{Donley2002}.  Follow-up observations of three of these candidates with {\it Chandra} a decade later, revealed that they had faded in the X-rays by factors of 240 to 6000, consistent with the expectations of the power-law decline of the accretion rate in a TDE \citep{Halpern2004}.  

The next X-ray survey with the wide-field coverage ($\sim$ 15 \% of sky), soft X-ray (0.2-2 keV) sensitivity capable of detecting TDEs was the XMM-Newton Slew Survey (XMMSL) starting in 2003, which discovered two galaxies with no detections in RASS or signs of an AGN in their optical spectra, that were detected as luminous soft X-ray sources in the XMMSL \citep{Esquej2007}, and subsequently shown to have faded in follow-up observations $2-3$ yr later by a factor consistent with a $t^{-5/3}$ power-law decline \citep{Esquej2008}.  However, one of the TDE candidates from XMMSL, NGC 3599, was found to have unanalyzed archival slew observations which showed that the galaxy was bright in the X-rays 18 months before the XMMSL detection, and thus did not have the flaring behavior of a bonafide TDE \citep{Saxton2015}.  More recently, the XMM-Newton slew survey has discovered five more candidates, one of which, XMMSL1 J0740-85, is notable for a well sampled light curve from follow-up {\it Swift}-XRT and pointed {\it XMM-Newton} observations \citep{Saxton2017}, a detection of a UV-bright transient component from {\it Swift}-UVOT follow-up, a detection in the radio from a non-relativistic outflow or a weak jet \citep{Alexander2017}, and a non-thermal $\Gamma \sim 2$ power-law component to the X-ray flare.


A couple more TDE candidates have been discovered by following up archival luminous and extremely soft ROSAT PSPC sources with {\it XMM-Newton} and/or {\it Chandra} \citep{Cappelluti2009, Maksym2014b}, including a candidate from a dwarf galaxy with a central black hole estimated to be $\lsim 2 \times 10^{6} M_\odot$ as inferred from the host galaxy luminosity \citep{Maksym2014b}.  A more recent cross match of sources that were detected by RASS but then had faded in serendipitous {\it XMM-Newton} observations taken two decades later, revealed three more sources broadly consistent with a TDE, potentially implying a TDE rate of $\sim 3 \times 10^{-5}$ yr$^{-1}$ per galaxy, but with no optical spectroscopic follow-up observations to investigate the nature of their host galaxies \citep{Khabibullin2014}.  

Several soft X-ray outbursts were also detected from archival searches of {\it Chandra} data, in particular of rich galaxy clusters with multiple epochs of observations, including a TDE candidate in a dwarf galaxy WINGS J1348+26 in cluster Abell 1795 \citep{Maksym2013}, with a host galaxy stellar stellar mass indicating a black hole in the intermediate-mass black hole range, of $1-5 \times 10^{5} M_\odot$ \citep{Maksym2014}, and a serendipitous detection of the flare in the EUV by the {\it Extreme Ultraviolet Explorer} ({\it EUVE}) satellite \citep{Donato2014}.  The large-amplitude, flaring soft X-ray source 3XMM J1521+07 was discovered in archival {\it Chandra} and {\it XMM-Newton} observations of a galaxy group NGC 5813, and identified as a TDE candidate, albeit with a very shallow decline rate post peak, argued to be the signature of slow circularization and super Eddington accretion \citep{Lin2015}.  

While all of the candidates from X-ray observations are convincing because of their extremely soft, thermal X-ray spectra, and large variability amplitudes unique from AGN, one of the weakness of these candidates is the lack of high cadence coverage of the X-ray light curves, and multiwavelength coverage to better constrain the broadband spectral energy distributions (SEDs) of the flares.  Fortunately, optical surveys have played an important role in enabling prompt discovery and follow-up, as well as a great improvement in time sampling of days instead of years.

\subsection{UV Candidates}


The first TDE candidate from UV observations was reported by \citet{Renzini1995} from archival {\it Hubble Space Telescope} ({\it HST)/FOC} observations of elliptical galaxy NGC 4552, which showed an enhanced nuclear flux in an epoch in 1993 compared to 1991.  However, {\it HST}/FOS follow-up spectroscopy by \citet{Cappellari1999} found broad-line emission from the nucleus, indicative of a low-luminosity AGN.  


The first search for TDEs in UV survey data was done with the {\it Galaxy Evolution Explorer} ({\it GALEX}) Deep Imaging Survey, which surveyed a total of 80 deg$^{2}$ of sky in the $NUV$ and $FUV$ bands in order to produce deep UV images to map the star formation history of galaxies out to $z \sim 2$ \citep{Martin2005}.  A systematic search for extragalactic UV transients from creating yearly coadds of the multiple exposures taken to build up the {\it GALEX} DIS images, and selecting only host galaxies with follow-up optical spectra that indicated an inactive galaxy host, yielded three candidates, two of which, D1-9 and D3-13, were serendipitously detected as an optical transient in $g$, $r$, $i$, and $z$-band difference imaging from the CFHT Legacy Survey (CFHTLS) \citep{Gezari2006, Gezari2008, Gezari2009}.  The third \textsl{GALEX} candidate, D23H-1, was followed up with prompt optical imaging that detected a coincident optical transient \citep{Gezari2009}, however, this candidate is now suspect, since a significant level of variability was detected in the mid-infrared $2-7$ years after the peak from serendipitous {\it WISE} observations, which suggest that the event could be associated with persistent variability of an AGN \citep{vanVelzen2016}, which was masked by the strong emission lines powered by star-formation in the host galaxy spectrum \citep{Gezari2009}.  

The multi-band detections of these TDE candidates in the $FUV$, $NUV$, $g$, $r$, $i$, and $z$ bands were well fitted by thermal blackbody emission, although with blackbody temperatures an order of magnitude cooler than had been measured for the X-ray selected TDEs with $T_{\rm BB} \sim (5-6) \times 10^4$ K and $L_{\rm BB} \sim (4-6) \times 10^{43}$ erg s$^{-1}$ \citep{Gezari2009}.  {\it Chandra} follow-up X-ray observations of the {\it GALEX} TDE candidates, $1-2$ years after their UV/optical peak, revealed weak, extremely soft X-ray emission in D1-9 and D3-13 \citep{Gezari2008}, indicating a second, higher temperature component ($T_{\rm BB} \sim (3-5) \times 10^5$ K), distinct from the UV/optical thermal emission \citep{Gezari2009}.

While the UV light curves constructed from {\it GALEX} DIS images with a yearly cadence were consistent with a flaring event, the optical light curves extracted from CFHTLS with a much higher cadence of 4 days, were in remarkably good agreement with numerical simulations for the fallback rate expected in a TDE, although with evidence for a shallower decline rate at late times \citep{Gezari2009}.  Furthermore, the optical difference imaging from CFHTLS localized the position of the flares to within 0.25 arcsec of the host galaxy nucleus, a great improvement over the 5 arcsec PSF {\it GALEX}.  The detection of TDEs in optical survey data, designed for the detection of supernovae, had powerful implications for the next generation of optical surveys, such as Pan-STARRS and PTF, to detect large numbers of TDEs, and in real-time. 

\subsection{Optical Candidates}

Before the first real-time searches for TDEs from optical time domain data, \cite{vanVelzen2011} performed an archival study of the 300 deg$^2$ SDSS Stripe 82 survey with the 2.5m telescope at Apache Point Observatory multi-epoch data to search for nuclear transients in inactive galaxy hosts.  This search yielded two candidates, TDE1 and TDE2, which had serendipitous UV detections by \textsl{GALEX}, and again were well described by thermal emission with $\sim 2 \times 10^4$ K, and a light curve that could be fitted with a power-law decline, although shallower than the $t^{-5/3}$ decline expected for the fallback rate.  TDE2 had a serendipitous follow-up spectrum during the flare, which showed evidence of transient broad H$\alpha$ emission associated with the event.  One of the most important results of this study, was that the properties of the optical light curve, namely their blue optical color, and lack of color evolution with time, could be used to distinguish TDEs from more common interlopers like variable AGN and SNe.  This study also measured an empirical TDE rate of 1.9 yr$^{-1}$ ($\Omega$/300 deg$^{2}$) ($m_{\rm lim}$/22 mag)$^{-3/2}$, which could be scaled to the flux limit ($m_{\rm lim}$) and area ($\Omega$) of future optical surveys.

\begin{figure}[t]
\begin{subfigure}[t]{0.5\textwidth}
\centering
\includegraphics[width=\textwidth]{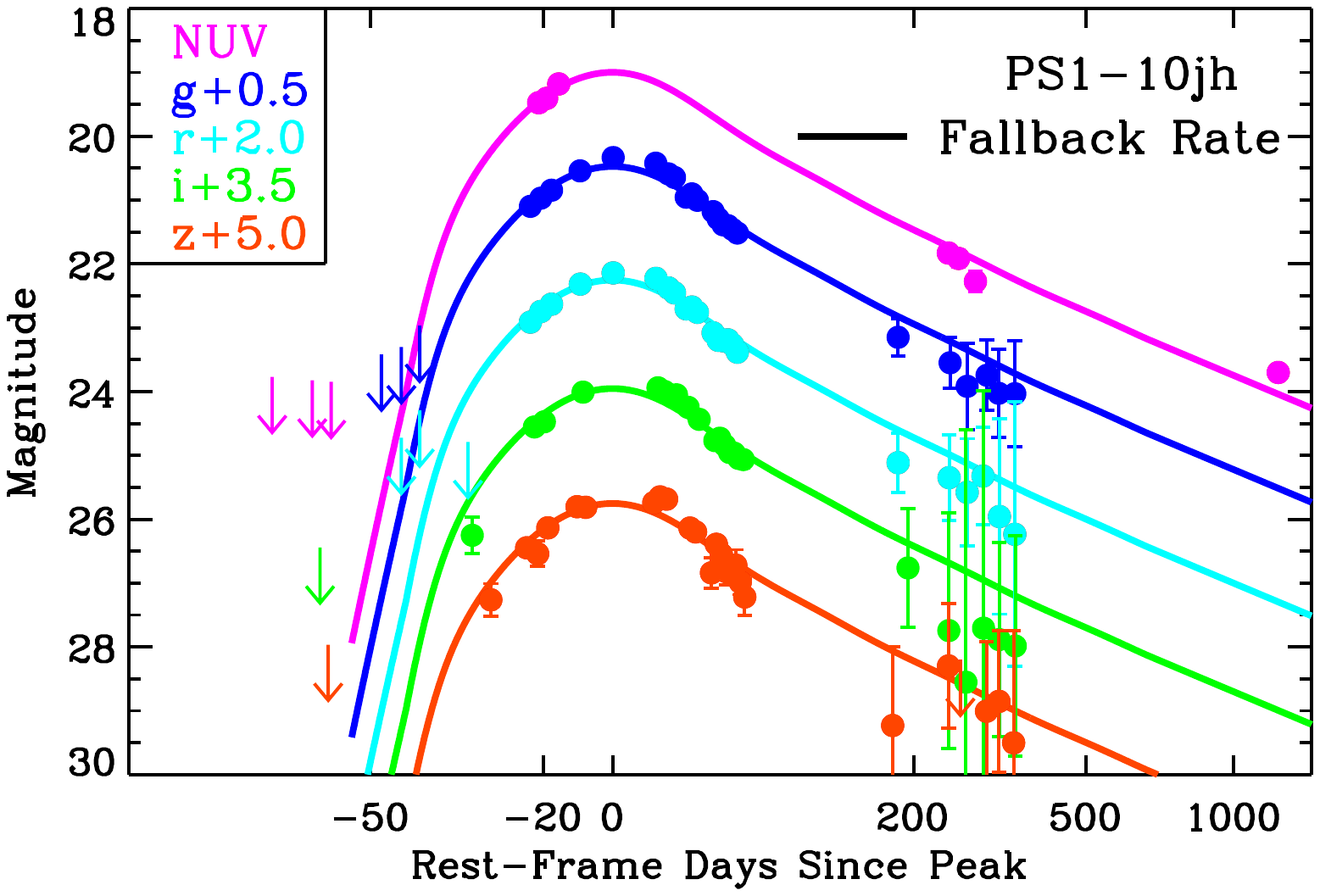}
\end{subfigure}
\begin{subfigure}[t]{0.4\textwidth}
\centering
\includegraphics[width=\textwidth,trim=0cm 0cm 9cm -8cm, clip]{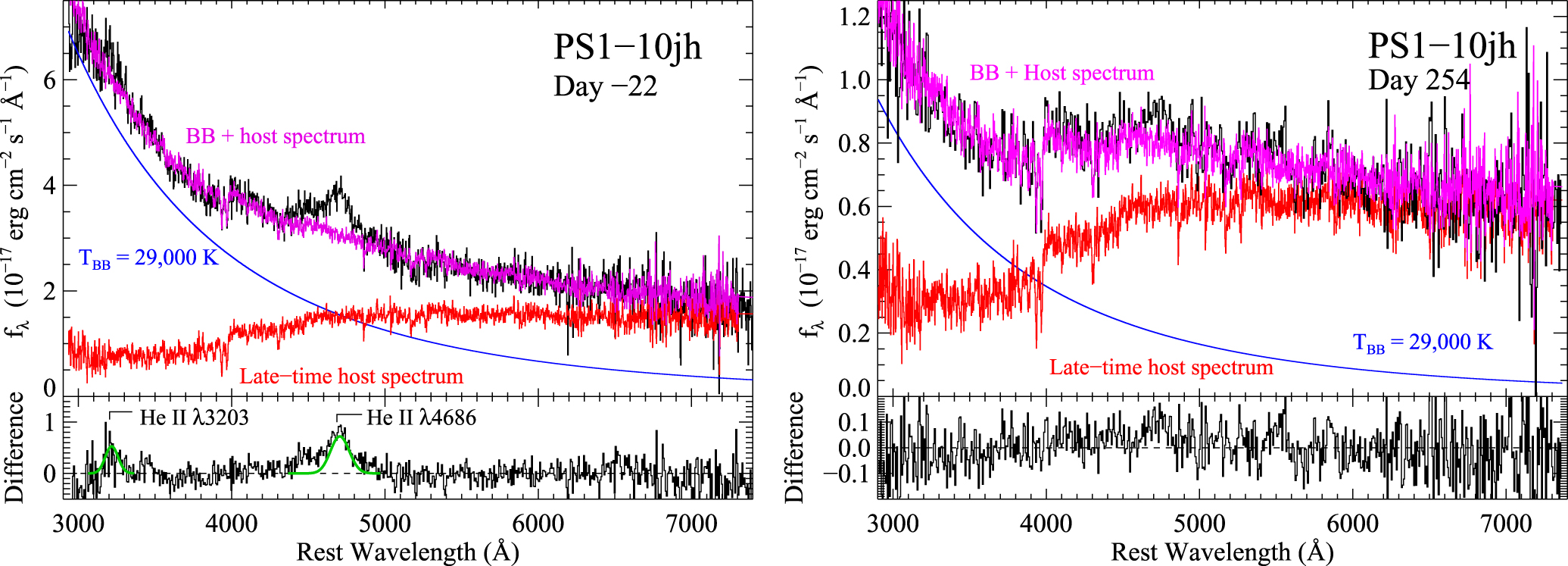}
\end{subfigure}
\caption{\footnotesize {\it Left}: {\it GALEX} Time Domain Survey $NUV$ and Pan-STARRS1 Medium Deep Survey optical $g$, $r$, $i$, and $z$ light curve of TDE PS1-10jh, with a late-time detection by HST/WFC3 in the $NUV$.  Lines show a numerical simulation for the fallback rate in a TDE from \cite{Guillochon2013} scaled to fit the light curve.  {\it Right}: Detection of transient, broad He II line emission on top of the hot, thermal continuum 22 days before the peak of the flare in TDE PS1-10jh.  Figures adapted from \cite{Gezari2015}.}
\label{fig:10jh}
\end{figure}

There was a huge jump in quality of TDE light curves with the search for TDEs in real time from optical time domain surveys.  PS1-10jh was the first TDE detected with a well-sampled rise to peak in optical survey data, and was discovered in a joint search by \cite{Gezari2012} for nuclear transients in the 2-day cadence {\it GALEX} Time Domain Survey in the $NUV$ and the 3-day cadence $g$,$r$,$i$, and $z$ band optical difference imaging of the 70 deg$^{2}$ Pan-STARRS1 (PS1) Medium Deep survey (MDS) taken with the 1.8 m telescope on Haleakala Observatory \citep{Chambers2016}.  Figure \ref{fig:10jh} shows the well-sampled {\it GALEX}+PS1 light curve of PS1-10jh, including a late-time $NUV$ detection from HST/WFC3 \citep{Gezari2015}, and a fit to a numerical simulation for the fallback rate in a tidal disruption event from \cite{Guillochon2013}.  The light curve is distinct from a SN for its remarkably constant blue optical color and persistent UV-bright emission, and it is surprising how well the numerical simulations describe the light curve shape, with no additional physics added to take into account accretion physics or radiative transfer effects.  

Another important aspect of this event was the first unambiguous detection of transient, broad emission lines in prompt optical spectroscopic follow-up observations.  Interestingly, the most prominent broad spectral feature in PS1-10jh was misidentified as Mg II $\lambda 2800$ at $z \sim 1$.  However, when the transient faded enough to reveal the underlying host galaxy absorption features 7.5 months after peak at $z = 0.1696$, the broad line in the peak spectrum was identified as He II $\lambda 4686$ with FWHM = 9,000 km s$^{-1}$, along with a detection of broad He II$\lambda 3203$, and no detection of broad Balmer line emission, which would be expected to have (see Figure \ref{fig:10jh}).  A careful subtraction of a late-time spectrum of the host galaxy placed further limits on the He II/H$\alpha$ ratio to be $> 5$ in the pre-peak spectrum \citep{Gezari2015}.  This first spectroscopic TDE was to be the first of a rare class of helium-only (TDE-He) TDEs.  

A second TDE candidate was discovered from search for blue, UV-bright nuclear transients in the {\it GALEX} and PS1 data, PS1-11af \citep{Chornock2014}, and was similar in UV/optical blackbody temperature and light curve shape as the other UV/optical TDEs.  This higher-redshift TDE, at $z=0.4046$, was notable for having a relatively featureless spectrum, except for transient broad absorption features in the UV at early times reminiscent of the O II "W" feature seen in some superluminous supernovae (SLSNe), but without the emergence of any optical line features or significant cooling, as seen in SLSNe.

Pan-STARRS has continued surveying the Northern Sky in the wide optical filter that covers the spans the $g,r,i$ bands as the Pan-STARRS Survey for Transients, and reported the discovery of AT2017eqx/PS17dhz, which had a typical UV/optical light curve of a TDE, but demonstrated the first case of a transformation in TDE spectral class, from a TDE-H+He to TDE-He spectrum, with initially broad H and He II lines, with the broad H lines disappearing on a timescale of $\sim 100$ days \citep{Nicholl2019}.  Note that this is the first TDE for which there is a "survey name" and an "astronomical transient (AT)" name, assigned by the Transient Name Server (TNS).  The TNS AT name is the official IAU mechanism for reporting new astronomical transients, and is important for keeping track of transients that are increasingly being detected by multiple surveys.  It is the convention to assign the "survey name" from the survey that first registered the coordinates of the transient with the TNS.

An archival analysis of transients in the Palomar Transient Factory (PTF) survey with the Palomar 48in telescope, with follow-up spectra and peak absolute magnitudes intermediate between SNe Ia and SLSNe, ($-19$ mag $ < M_R < -21$ mag) produced three nuclear transients, one of which, PTF 09ge, had a very similar light curve shape and He II only spectrum to that of PS1-10jh \citep{Arcavi2014}.  The other two, PTF09djl and PTF09axc, had less well-sampled light curves, but did show very broad Balmer line features, representing a class of H-rich (TDE-H) TDEs, and interestingly, had strong H$\delta$ absorption characteristic of the rare class of E+A galaxies, whose strong A-star features and no emission lines are explained as post-starburst galaxies with a truncation of star formation $\sim 1$ Gyr ago.  PTF had detected another interesting blue, nuclear transient, PTF10iya, in a star-forming galaxy \citep{Cenko2012b}, however, with its poorly sampled light curve, featureless optical spectrum, and non-thermal X-ray spectrum ($\Gamma \sim 2$) , the association of the short-lived ($\sim 10$ d) event with a TDE was not clear-cut.

A real-time search for TDEs was conducted during the next phase of PTF operations, known as iPTF.  During this search, the strategy for selecting TDEs was to use complete {\it Neil Gehrels Swift} UV follow-up of all nuclear transients to filter out AGN and SNe (Figure \ref{fig:sdss}).  This systematic UV follow-up program, which due to finite {\it Swift} observing time was limited to nuclear transients in red galaxy hosts (characteristic of previous TDE discoveries), yielded three more nuclear transients with properties consistent with previously discovered optical TDEs: iPTF-16axa \citep{Hung2017}, iPTF-16fnl \citep{Blagorodnova2017}, and iPTF-15af \citep{Blagorodnova2019}.    At the time, iPTF-16fnl was the closest optical TDE ever detected, in a nearby, E+A galaxy at $z=0.0163$, and with a lower peak luminosity ($L_{\rm BB} \sim 10^{43}$ erg s$^{-1}$) and faster fading timescale than previous TDEs, with important implications for the TDE luminosity function and rate.  An {{\it HST}/STIS UV spectrum of iPTF-15af revealed broad absorption and emission line features analogous to broad absorption-line quasars (BAL QSOs), suggesting the presence of an outflow, and discussed further in \S \ref{sec:outflow}.


One of the unexpected results from the All-Sky Automated Survey for Supernovae (ASAS-SN) survey \citep{Kochanek2017}, a network of 14 cm telescopes hosted by the Las Cumbres Observatory, is that despite its large pixel size (8 arcsec) compared to other optical surveys, it would be highly efficient in detected tidal disruption events in the nuclei of galaxies.  One of the reasons for this is the bright limit of ASAS-SN ($m < 17$ mag), and manageable detection rate, allowing for near spectroscopic completeness  \citep{Brown2019}, and thus no selection bias in spectroscopically following-up and classifying transients.  In the first two years of the ASAS-SN survey, they discovered four TDEs: ASASSN-14ae \citep{Holoien2014}, ASASSN-14li \citep{Holoien2016}, ASASSN-15lh \citep{Dong2016, Leloudas2016}, and ASASSN-15oi \citep{Holoien2016b}.  The bright, nearby TDEs discovered by ASAS-SN allow for high signal-to-noise multi-wavelength follow-up and spectroscopic monitoring, and we will discuss each of these important TDE discoveries as we discuss the exciting results from their multiwavelength observations, including detections in the X-ray, mid-infrared, radio.  

In the last two years, ASAS-SN has discovered another five TDEs (AT2018fyk/ASASSN-18ul, AT2018dyb/ASASSN-18pg, AT2018hyz/ASASSN-18zj, AT2019ahk/ASASSN-19bt, and AT2019azh/ASASSN-19dj), an impressive discovery rate that is now only recently being outpaced by the ZTF survey.  One of these recent TDE discoveries, AT2019ahk/ASASSN-19bt, was detected in the {\it Transiting Exoplanet Survey Satellite} ({\it TESS}) Continuous Viewing Zone (CVZ), which enabled the extraction of a light curve in the {\it TESS} broadband optical filter binned into a cadence of just 2 hr over a timescale of 2 months \citep{Holoien2019b}!  This allowed for an extremely detailed look at the pre-peak light curve of the TDE, which was surprisingly smooth, and well fitted to a $\propto t^2$ rise to peak in the first 15 days of its 41 day rise to peak.


The Zwicky Transient Facility \citep[ZTF]{Bellm2019}, which uses an upgraded detector on the Palomar 48in telescope with a 50 deg$^{2}$ field of view and a limiting magnitude per epoch of $m_{\rm lim} \sim 20.5$ mag in the $g$ and $r$ bands, as well as a dedicated robotically operated low-resolution ($R \sim 100$) SED Machine (SEDM) spectrograph on the Palomar 60in telescope \citep{Blagorodnova2018} for classification of transients brighter than $19$ mag, has yielded the largest sample of TDEs from a single survey to date.  In the first 1.5 yr of survey operations, they reported the systematic selection, classification, and characterization of 17 TDEs \citep{vanVelzen2020}, some of which were detected or discovered by other surveys, including ASAS-SN, the Asteroid Terrestrial-impact Last Alert System \citep[ATLAS]{Tonry2018}, and the Pan-STARRS Survey for Transients \citep{Nicholl2019}. 

\subsection{Hard X-ray Candidates}
There was a big shift in the TDE landscape on 2011 March 25, when {\it Swift}/BAT discovered a long-duration hard X-ray transient, accompanied by transient radio and infrared emission, from the nucleus of a quiescent galaxy at $z=0.3543$: Swift J1644+57 \citep{Bloom2011, Levan2011, Zauderer2011}.  The hard, power-law X-ray spectrum, highly super Eddington luminosity (isotropic X-ray luminosity of $\sim 10^{47}$ erg s$^{-1}$), radio synchrotron spectrum, and power-law decline of the X-ray light curve were explained as viewing an on-axis collimated jet launched by the tidal disruption of a star by the galaxy's central $\sim 10^6 M_\odot$ black hole.  In fact, a class of "jetted-TDEs" had actually been predicted by \cite{Giannios2011}, but their models focused on off-axis emission from the interaction of the jet with the interstellar medium, and radio searches had only yielded a few late-time detections \citep{Bower2011, Bower2013}, implying that $\sim 10\%$ of TDEs launch jets.  Two more jetted TDE candidates were detected by {\it Swift} with similar X-ray and radio properties: Swift J2058+05 \citep{Cenko2012, Pasham2015} and Swift J1112-82 \citep{Brown2015}.  The implyed rate of jetted TDEs is only $\sim 3 \times 10^{-10}$ yr$^{-1}$ per galaxy, suggesting a combination of the effects of a small beaming angle ($\sim 1 \deg$), and that only $\sim 10\%$ of TDEs produce relativistic jets \citep{Brown2015}. 

 In the case of Swift J2015+05 a thermal component ($T_{\rm BB} \sim 3 \times 10^4$ K) in the UV and optical was detected, similar to the UV and optically selected TDEs, and which may have been missed in Swift J1644+57 due to dust extinction in the host galaxy nucleus.  A systematic search of 53,000 galaxies within 100 Mpc with observations in the {\it Swift}/BAT archive yielded the association of a hard X-ray flare from nine quiescent galaxies, implying a candidate TDE rate in the hard X-rays of $2 \times 10^{-5}$ yr$^{-1}$ per galaxy, a rate comparable to that measured for TDEs selected in the soft X-rays, suggesting that hard X-ray components could be ubiquitous even in unbeamed TDEs \citep{Hyrniewicz2016}.  However,  with sparsely sampled X-ray light curves, and no multiwavelength follow-up data, it is hard to verify their classification as bonafide TDEs.

{\it Integral} discovered a flaring hard X-ray source in the 20-40 keV band, which was also detected in the 2-10 keV band in follow-up observations with {\it Swift}/XRT and {\it XMM-Newton}, from the nucleus of the nearby spiral galaxy NGC 4845, which was argued from X-ray spectrum, luminosity and timing arguments, to be emission from an accretion disk corona formed from the tidal disruption of a super-Jupiter sized object around a $\sim 3 \times 10^5 M_\odot$ black hole \citep{Nikolajuk2013} and was also detected as a radio transient consistent with emission from an off-axis jet \citep{Lei2016}.  The tidal disruption of a substellar object was also invoked to explain the low peak luminosity of the ROSAT soft X-ray selected TDE in NGC 5905 \citep{Li2002}.

\begin{table}[h]
\tabcolsep7.5pt
\caption{Table of tidal disruption events reported in the literature.}
\label{tab1}
\begin{center}
\begin{tabular}{@{}l|l|c|l|c|c|l@{}}
\hline
Name & Survey$^{\rm a}$ & Waveband & Redshift & log L$_{\rm BB}^{\rm b}$ & log T$_{\rm BB}$ & Publication$^{\rm c}$\\
 & & & & (erg s$^{-1}$) & (K) & \\
\hline
NGC 5905 & ROSAT & X & 0.01124 & 40.94 & 5.84 & \cite{Bade1996} \\
RX J1624+75 & ROSAT & X & 0.0636 & 43.38 & 6.05 & \cite{Grupe1999} \\
RX J1242-11A & ROSAT & X & 0.050 & 42.60 & 5.84 & \cite{Komossa1999} \\
RX J1420+53 & ROSAT & X & 0.147 & 43.38 & 5.67 & \cite{Greiner2000} \\
GALEX D3-13 & GALEX & U & 0.3698 & 43.98 & 4.66 & \cite{Gezari2006} \\
SDSS J1323+48 & XMM & X & 0.0875 & 44.30 & 5.91 & \cite{Esquej2007} \\
GALEX D1-9 & GALEX & U & 0.326 & 43.48 & 4.59 & \cite{Gezari2008} \\
TDXF J1347-32 & ROSAT & X & 0.0366 & 42.73 & 6.14 & \cite{Cappelluti2009} \\
GALEX D23H-1 & GALEX & U & 0.1855 & 43.95 & 4.70 & \cite{Gezari2009} \\
SDSS J1311-01 & Chandra & X & 0.195 & 41.74 & 6.14 & \cite{Maksym2010} \\
SwiftJ1644 & Swift & G & 0.353 &   n/a & n/a & \cite{Bloom2011} \\
2XMMi J1847-63 & XMM & X & 0.0353 & 42.82 & 5.96 & \cite{Lin2011} \\
SDSS-TDE1 & SDSS & O & 0.136 & 43.64 & 4.42 & \cite{vanVelzen2011} \\
SDSS-TDE2 & SDSS & O &      0.2515 & 44.54 & 4.37 & \cite{vanVelzen2011} \\
PS1-10jh & PS & O &      0.1696 & 44.47 & 4.59 & \cite{Gezari2012} \\
SDSS J1201+30 & XMM & X & 0.146 & 45.00 & 6.06 & \cite{Saxton2012} \\
SwiftJ2058 & Swift & G & 1.186 &   n/a & n/a & \cite{Cenko2012} \\
WINGS J1348+26 & Chandra & X & 0.0651 & 41.79 & 6.06 & \cite{Maksym2013} \\
PS1-11af & PS & O &      0.4046 & 44.16 & 4.28 & \cite{Chornock2014} \\
RBS 1032 & ROSAT & X & 0.026 & 41.70 & 6.11 & \cite{Maksym2014} \\
PTF-09ge & PTF & O &     0.064 & 44.04 & 4.08 & \cite{Arcavi2014} \\
PTF-09axc & PTF & O &      0.1146 & 43.46 & 4.08 & \cite{Arcavi2014} \\
PTF-09djl & PTF & O &      0.184 & 44.42 & 4.41 & \cite{Arcavi2014} \\
ASASSN-14ae & ASASSN & O &     0.0436 & 43.87 & 4.29 & \cite{Holoien2014} \\
3XMM J1521+07 & XMM & X & 0.179 & 43.51 & 6.30 & \cite{Lin2015} \\
SwiftJ1112 & Swift & G & 0.89 &   n/a & n/a & \cite{Brown2015} \\
ASASSN-14li & ASASSN & O &     0.02058 & 43.66 & 4.52 & \cite{Holoien2016} \\
ASASSN-15lh & ASASSN & O & 0.2326 & 45.34 & 4.30 & \cite{Dong2016} \\
ASASSN-15oi & ASASSN & O &       0.0484 & 44.45 & 4.60 & \cite{Holoien2016} \\
iPTF-16axa & PTF & O &      0.108 & 43.82 & 4.46 & \cite{Hung2017} \\
iPTF-16fnl & PTF & O &     0.0163 & 43.18 & 4.47 & \cite{Blagorodnova2017} \\
3XMM J1500+01 & XMM & X & 0.1454 & 43.08 & 6.06 & \cite{Lin2017} \\
OGLE16aaa & OGLE & O &      0.1655 & 44.22 & 4.36 & \cite{Wyrzykowski2017} \\
XMMSL1 J0740-85 & XMM & X & 0.0173 & 42.61 & 6.00 & \cite{Saxton2017} \\
iPTF-15af & PTF & O &     0.07897 & 44.10 & 4.85 & \cite{Blagorodnova2019} \\
AT2017eqx/PS17dhz & PS & O &      0.1089 & 43.82 & 4.32 & \cite{Nicholl2019} \\
AT2018zr/PS18kh & PS & O &     0.071 & 43.78 & 4.14 & \cite{vanVelzen2020} \\
AT2018bsi/ZTF18aahqkbt & ZTF & O &     0.051 & 43.87 & 4.53 & \cite{vanVelzen2020} \\
AT2018dyb/ASASSN-18pg & ASASSN & O &     0.018 & 44.08 & 4.40 & \cite{Leloudas2019} \\
AT2018fyk/ASASSN-18ul & ASASSN & O &     0.059 & 44.48 & 4.54 & \cite{Wevers2019} \\
AT2018hco/ATLAS18way & ATLAS & O &     0.088 & 44.25 & 4.39 & \cite{vanVelzen2020} \\
AT2018hyz/ASASSN-18zj & ASASSN & O &     0.04573 & 44.10 & 4.25 & \cite{Gomez2020} \\
AT2018iih/ATLAS18yzs & ATLAS & O &      0.212 & 44.62 & 4.23 & \cite{vanVelzen2020} \\
AT2018lna/ZTF19aabbnzo & ZTF & O &     0.091 & 44.56 & 4.59 & \cite{vanVelzen2020} \\
AT2018lni/ZTF18actaqdw & ZTF & O &      0.138 & 44.21 & 4.38 & \cite{vanVelzen2020} \\
AT2019ahk/ASASSN-19bt & ASASSN & O &     0.0262 & 44.08 & 4.30 & \cite{Holoien2019} \\
AT2019azh/ASASSN-19dj & ASASSN & O &     0.0222 & 44.50 & 4.51 & \cite{vanVelzen2020} \\
AT2019bhf/ZTF19aakswrb & ZTF & O &      0.1206 & 43.91 & 4.27 & \cite{vanVelzen2020} \\
AT2019cho/ZTF19aakiwze & ZTF & O &      0.193 & 43.98 & 4.19 & \cite{vanVelzen2020} \\
AT2019dsg/ZTF19aapreis & ZTF & O &       0.0512 & 44.26 & 4.59 & \cite{vanVelzen2020} \\
AT2019ehz/Gaia19bpt & Gaia & O &     0.074 & 44.03 & 4.34 & \cite{vanVelzen2020} \\
AT2019eve/ZTF19aatylnl & ZTF & O &     0.0813 & 43.14 & 4.06 & \cite{vanVelzen2020} \\
AT2019lwu/ZTF19abidbya & ZTF & O &      0.117 & 43.60 & 4.14 & \cite{vanVelzen2020} \\
AT2019meg/ZTF19abhhjcc & ZTF & O &      0.152 & 44.36 & 4.44 & \cite{vanVelzen2020} \\
AT2019mha/ATLAS19qqu & ATLAS & O &      0.148 & 44.05 & 4.35 & \cite{vanVelzen2020} \\
AT2019qiz/ZTF19abzrhgq & ZTF & O &     0.0151 & 43.44 & 4.27 & \cite{vanVelzen2020} \\
\hline
\end{tabular}
\end{center}
\begin{tabnote}
$^{\rm a}$Survey that first discovered the nuclear transient; $^{\rm b}$Integrated blackbody luminosity except for X-ray selected TDEs for which the absorbed luminosity in 0.3-2 keV band is given; $^{\rm b}$ Publication in which the luminosity and temperature were used.
\end{tabnote}
\end{table}
 \label{table1}

\section{OBSERVED PROPERTIES}

It is now possible, with the growing census of observed TDEs listed above, to compare them in detail to the theoretical expectations for emission from TDEs, and attempt to extract the physical parameters of the events, including black hole mass and stellar type.  As we will see in the following sections, however, there are many discrepancies between the observations and the basic predictions for emission from an accreting debris disk, requiring important modifications to the simplest emission models.

\subsection{Light Curves}  \label{sec:time}

The tidal disruption of a star occurs when it approaches the distance from the black hole at which the tidal forces exceed the self gravity of the star, 

\begin{equation}
 \frac{GMR_\star}{r^3} > G\frac{M_\star}{R_\star^2}
\end{equation}

\noindent resulting in a tidal disruption radius, 

\begin{equation}
R_{\rm T} = R_\star \left (\eta^2 \frac{M_{\rm BH}}{M_\star} \right )^{1/3},
\end{equation}

\noindent where $\eta \simeq 1$, and depends weakly on the structure of the star \citep{Phinney1989, Evans1989}.  After disruption, the stellar debris will have a spread of specific binding energy, 

\begin{equation}
\Delta \epsilon = \pm \frac{GM_\star}{R_{\rm T}^2} R_\star = \frac{GM_\star}{R_\star} \left (\frac{M_{\rm BH}}{M_\star} \right )^{1/3}
\end{equation}

\noindent such that half the mass will be gravitationally bound to the black hole ($\Delta \epsilon < 0$) and thus available to be accreted.  The fallback timescale ($t_{\rm fb}$), the characteristic minimum timescale in a TDE, is defined as the orbital period of the most bound debris, 

\begin{equation}
t_{\rm fb} = 2\pi GM_{\rm BH} (2E)^{-3/2} = \frac{\pi}{M_\star} \left (\frac{M_{\rm BH} R_\star^3}{2G} \right) ^{1/2} = 0.11 {\rm yr}~r_\star^{3/2} M_6^{1/2} m_\star^{-1}
\end{equation}

The fact that $t_{\rm fb}$ scales with the square-root of the black hole mass, implies that the timing of TDE flares should in principle be used to yield information on the mass of the central black hole.  Another fundamental property of the fallback of the debris streams, is that if their specific energy distribution is uniform, i.e. dE/dM = 0, then the rate at which material returns to pericenter can be derived as:

\begin{equation}
\frac{dM}{dE}\frac{dE}{dt} = \frac{2\pi}{3} (GM_{\rm BH})^{2/3} \frac{dM}{dE}t^{-5/3} 
\end{equation}

\noindent and follows a power-law, $dM/dt \propto (t-t_{\rm D})^{-5/3}$.  In the case of a {\it partial} disruption, a steeper power-law decline is expected, $dM/dt \propto (t-t_{\rm D})^{-9/4}$ \citep{Coughlin2019}.  In addition, the internal structure \citep{Ramirez-Ruiz2009, Guillochon2013, Golightly2019} and spin \citep{Golightly2019b} of the star, as well as the spin of the black hole \citep{Kesden2012, Gafton2019} and the impact parameter of the star's orbit \citep{Gafton2019}, will have an imprint on the energy distribution of the debris, and thus the fallback rate.  

One of the most remarkable observed characteristics of TDEs is that at face value, they do appear to have a light curve that follows the general shape of the theoretical TDE fallback rate.  In fact, when one fits a $t^{-5/3}$ power law to the light curve on its decline from peak, there is a strong correlation with the time of peak since the inferred time of disruption, $\Delta t = (t_{\rm peak}-t_{\rm D}$), and the estimated black hole mass, where $\Delta t \propto M_{\rm BH}^{1/2}$, as would be expected for the fallback timescale \citep{vanVelzen2019, vanVelzen2020b}. 
 
However, this good agreement between the optical light curves and the $t^{-5/3}$ power-law of the fallback accretion rate is actually not entirely consistent with the theory.  For emission from a geometrically thin, and optically thick accretion disk, one would expect a gradual cooling over time with the declining accretion rate, since $T_{\rm disk} \propto \dot{M}^{1/4}$.  Thus for optical emission, which is on the Rayleigh Jeans tail of the hot blackbody emission characteristic of TDEs, $\nu L_\nu \propto T \propto \dot{M}^{1/4} \propto t^{-5/12}$ \citep{Lodato2011}, and one expects a {\it shallower} power-law in the optical bands.  
Furthermore, there does not appear to be a strong correlation between the rise-time to peak with black hole mass \citep{vanVelzen2020}, as one would expect if it was following the fallback rate.   Regardless, the fact that the observed light curve decline follows the fallback rate is still somewhat of a surprise, given that this assumes that the accretion rate is equivalent to the fallback rate, which allows for no time delay for the debris streams to shock and circularize into an accretion disk.  The physical implication is that either that accretion does proceed promptly, or that we are seeing emission from a process not related to the accretion disk, potentially from the debris stream collisions themselves.  One of the important ways to distinguish between these scenarios is by looking at the evolution of the soft X-ray emission in these events, which are a more direct tracer of the accretion flow near the central black hole, and are discussed in more detail in \S \ref{sec:opt_xray}.

\subsection{Peak Luminosities and Characteristic Temperatures}

The peak fallback rate, $\dot{M}_{\rm peak}  \sim \frac{1}{3} \frac{M_\star}{t_{\rm fb}}$, will result in a range of Eddington ratios, depending on the central black hole mass and the fraction of gas expelled in an outflow ($f_{\rm out}$),

\begin{equation}
\frac{ \dot{M} (t_{\rm fb}) }{\dot{M_{\rm Edd}}} = 133 (\eta/0.1) M_6^{-3/2} m_\star^{4/5}  (1-f_{\rm out}),
\end{equation} \label{eq:mdot}

\noindent where $L_{\rm Edd} = \eta \dot{M}_{\rm Edd} c^2 = 1.3 \times 10^{44} M_6$ erg s$^{-1}$.
This presents the possibility that the ``mode'' of the accretion flow will be vastly different for black holes of different mass, with black holes less than $3 \times 10^{7} M_\odot$ in the super-Eddington regime, and above $3 \times 10^{7} M_\odot$ in the sub-Eddington regime \citep{Strubbe2009, Lodato2011, Metzger2016}.  This Eddington limit should  imprint itself on the TDE luminosity function, if non-axisymmetric effects are not dominant.  
The estimation of the Eddington ratio requires a measurement of the central black hole mass.  For the two dozen TDEs with host galaxy velocity dispersion measurements used to estimate $M_{\rm BH}$, the inferred Eddington ratios appear to be clustered between $\sim 0.1-1$ for the UV/optical component, and with a much larger range for the soft X-ray component, with $\sim 10^{-3} - 1$, during peak \citep{Wevers2019}.  However, given the dramatic variability observed in X-rays from TDEs, as discussed in \S \ref{sec:xray}, there could be many factors contributing to the low Eddington ratios of the X-ray component in some cases, including due to the presence of obscuration, delayed accretion, and non-isotropic emission.

By combining the observed luminosities, with the flux limits of the individual surveys, \cite{vanVelzen2018} calculated a luminosity function for TDEs fitted by power-law, $dN/dL \propto L^{-2.5}$, with no correlation with black hole mass, as would be expected for Eddington-limited accretion.  However, \cite{vanVelzen2018} explain this as a potential result of the sensitivity of $\dot{M}_{fb}$ to $M_\star$.  Since $\dot{M}_{fb} \propto M_\star^{4/5}$ (see Equation \ref{eq:mdot}), then for an initial mass function of stars, where $dN_\star/dM_\star \propto M_\star^{-\alpha}$, where $\alpha \sim 2.3$, then one would expect a steep decrease of the event rate with peak luminosity.  However, there may be other effects on the peak luminosity, such as photon trapping in a radiatively driven wind, which we will address in Section \ref{sec:bh}.

\subsection{Spectral Classes}

The first definitive detection of transient line emission from a TDE was reported for PS1-10jh \citep{Gezari2012}.  This TDE was selected as a UV-bright, optically blue nuclear flare in a galaxy with an unknown redshift.  It's spectrum, taken 3 weeks before its peak actually delayed the classification of the event as a TDE, because on top of the very blue continuum, only a single strong broad emission line was detected, and assumed to be Mg II $\lambda$2798, a line seen in isolation in the optical spectra of higher redshift quasars ($1 \lsim z \lsim 1.5$) for which the H$\beta$ line has been redshifted out of the optical band.  It was not until the flare faded enough for the absorption features of the host galaxy to emerge at $z=0.1696$, that the broad-line detected in the spectrum was identified as He II $\lambda$4686, a line actually predicted by \citet{Ulmer1999} to be a signature of the hot photoionizing continuum in a TDE.  In fact, transient, broad He II emission had been detected in a Seyfert galaxy NGC 5548, and interpreted as the accretion of new material at small distances from the black hole, potentially from the tidal disruption of a star \citep{Peterson1986}.  

While the detection of He II$\lambda 4686$ fit into the expectations of a TDE spectrum, the complete lack of any hydrogen features was very surprising, but also confirmed that the gas being photoionized by the event was not ambient gas from the interstellar medium (ISM), but originated from the stellar debris itself.  The high He II$\lambda 4686$/H$\alpha$ ratio was explained either a as a signature of helium-rich gas from the tidal disruption of a stripped star \citep{Gezari2012, Bogdanovic2014}, or the result of suppression of hydrogen line emission due to optically-thick reprocessing \citep{Roth2016}.  Soon after, archival detections of three new optical TDE candidates were reported by the Palomar Transient Factory (PTF) survey, the one for which they had the best sampled light curve, PTF09ge, also showed only broad He II line emission in its spectrum \citep{Arcavi2014}!  However, two of the PTF TDE candidates showed only broad H Balmer-line emission, and the first TDE from the ASAS-SN survey, ASASSN-14ae \citep{Holoien2014} showed both H and He II features, leading \citet{Arcavi2014} to suggest a continuum of spectral types, from H-dominated to intermediate H+He to He-dominated spectra.  

\begin{figure}[t]
\includegraphics[width=3in,trim=1cm 0 4cm 14cm, clip]{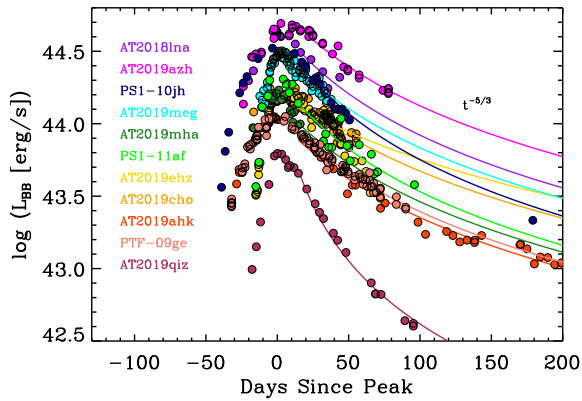}
\caption{\footnotesize Compilation of bolometric luminosity curves of TDEs with well-sampled pre-peak optical light curves, labeled by their AT name, with the exception of PS1-10jh, PS1-11af, and PTF-09ge.  The light curves were constructed by scaling the $r$-band light curve by the peak bolometric luminosity determined from a blackbody fit to the optical + UV photometry reported in \cite{vanVelzen2020b}, and assuming no evolution in temperature.  In the case of PTF-09ge no UV observations were taken at the time of the event, and so the bolometric luminosity is estimated from its optical spectrum.  In the case of AT2019ahk/ASASSN-19bt, I plot the {\it Swift} $uvw2$ light curve scaled by the peak bolometric luminosity.  Also shown is a $t^{-5/3}$ power-law decline fit to these curves after peak.  }
\label{fig:nicholl}
\end{figure}

\begin{figure}[t]
\begin{subfigure}[t]{0.5\textwidth}
\centering
\includegraphics[width=\textwidth,trim=2.cm 0 2.cm 0]{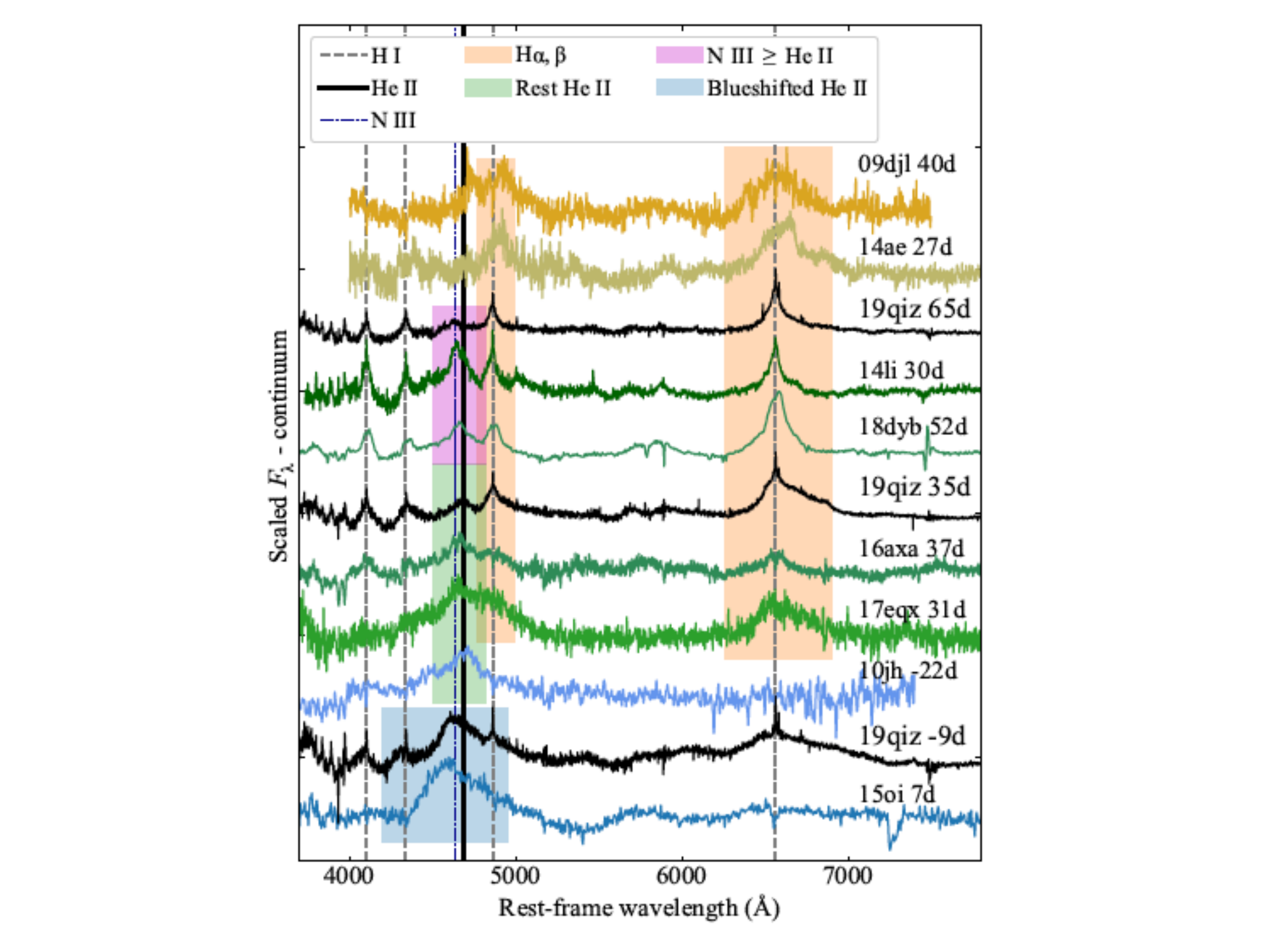}
\end{subfigure}
\begin{subfigure}[t]{0.45\textwidth}
\centering
\includegraphics[width=\textwidth,trim=1cm 0cm 6cm 15cm]{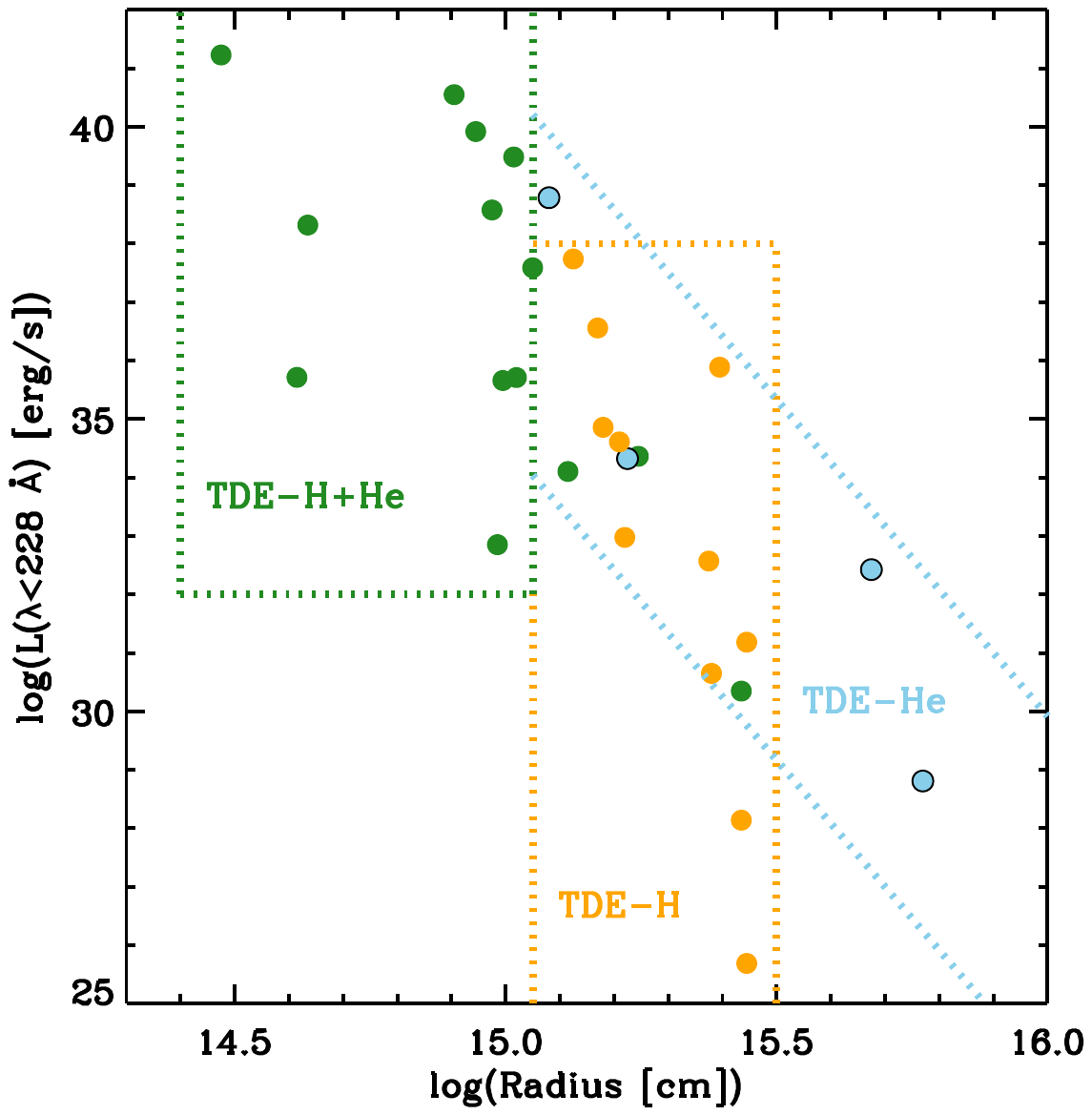}
\end{subfigure}
\caption{\footnotesize {\it Left:} Spectroscopic classes of TDEs labeled by their phase in days since peak, and highlighted by the presence of broad He II $\lambda 4686$ (TDE-He), the presence of broad He II and H$\alpha$ and H$\beta$ (TDE-H+He), the presence of strong Bowen fluorescence feature N III $\lambda 4640$, and those with only broad H features (TDE-H).  Figure from \cite{Nicholl2020}.  {\it Right}: Plot of the blackbody luminosity below 228 \AA capable of ionizing He and triggering the Bowen fluorescence mechanism, based on the temperature and bolometric luminosity of the UV/optical component, versus the inferred radius of the UV/optical photosphere.  Points are color-coded by their spectral class.  Note that TDE-H+He class appear to prefer more UV luminous flares with more compact radii than the TDE-H class.  TDE-He class appear in a larger range of radii and luminosities.}
\label{fig:spec}
\end{figure}

This continuum of spectral classes became more complex with the identification of Bowen fluorescence metal lines of O III$\lambda$3760 and N III $\lambda$ 4100 and 4640 in the "intermediate" H+He spectral type-TDEs iPTF15af \citep{Blagorodnova2019}, and in ASASSN-14li, iPTF16axa, and AT2018dyb/ASASSN-18bg \citep{Leloudas2019, Holoien2020}.  Interestingly, despite these lines being expected in systems with strong UV radiation ($\lambda < 228$\AA), Bowen flourescence lines had only recently been detected in AGN spectra in a new class of flaring AGN by \citet{Trakhtenbrot2019}.  This trend was further strengthened by the detection of these O III/N III features in all of the H+He TDEs from the ZTF survey (7 classified as TDE-H+He), with the remaining TDEs having either only broad H features (9 classified as TDE-H) or broad He II features (1 classified as TDE-He) \citep{vanVelzen2020}.  Indeed, He II$\lambda$4686 is expected to accompany Bowen lines, since the Bowen fluorescence mechanism is triggered by the ionization and recombination of He II \citep{Leloudas2019}, which excites transitions in N and O with similar energies to that of the He II Ly$\alpha$ photons \citep{Bowen1935}.  However, \citet{Leloudas2019} note that not all TDEs with strong He II lines exhibit the Bowen fluorescence lines, suggesting that they do not all have the optimal physical conditions for strong resonance to happen.  However, in fact, in the ZTF TDE sample, all TDEs with H and He II lines also show Bowen features, while it is only the spectra with He II only spectra that do not have Bowen features.

\cite{vanVelzen2020} presented a suggestive trend that spectral type correlates with the radius and temperature of the UV/optical photosphere, with TDE-Bowen spectra coming from smaller radii (i.e. a more compact region) and higher temperatures than the H-rich spectra, both physical conditions more conducive to powering Bowen fluorescence.  In Figure \ref{fig:spec} I plot the luminosity below 228 \AA~vs. radius for the optically selected TDEs as a function of their spectral class.  It is clear that the TDE-H+He class (including Bowen line features) prefer higher EUV luminosities and compact radii compared to the TDE-H class, while the TDE-He class show a range of conditions.   Another important clue is the relative rates of the spectral types.  Note that among the 17 TDEs in the Year 1.5 ZTF TDE sample, only one was He-only!  Given that the galaxy host properties and flare properties do not stand out for this one object, AT2018iih/ATLAS18yzs, it may be indicating that it was the disrupted star's properties that were unique (i.e. He-rich).  However, a simple picture connecting spectral class with TDE parameters is complicated further by the detection of a TDE by Pan-STARRS, AT2017eqx/PS17dhz, that transitioned over a timescale of 100 days from H-dominated to He-dominated, with the He II line blueshifted by 5000 km s$^{-1}$ \citep{Nicholl2019}.  \citet{Nicholl2019} explain this spectral transformation as due to a contracting atmosphere, eventually revealing a wind in the polar direction.  The contracting atmosphere is needed to suppress H emission at late times, which is expected in the \citet{Roth2016} model for optically thick atmosphere's with a certain radius.

In \cite{Wevers2019}, they noted the emergence of Fe II lines in AT2018fyk/ASASSN-18ul and ASASSN-15oi, signaling the presence of low-ionization lines from of optically thick, dense gas.  Furthermore, they argued that since the Fe II complex is visible during a secondary maximum in the UV/optical light curve of AT2018fyk/ASASSN-18ul, perhaps both the low-ionization lines and the UV/optical plateau are a result of reprocessing of the soft X-ray radiation into the UV/optical wavebands.

\subsection{Line Profiles}
The profiles of TDE emission lines are very broad ($(3-13) \times 10^{4}$ km s$^{-1}$ \citep{Arcavi2014, Hung2017, Holoien2019b}, and in some cases with double-peaked, boxy, and/or blue-shifted structures.
Such large velocities are expected in a TDE.  If one assumes that the lines are broadened by Doppler broadening of the gas in the gravitational potential of the central black hole, then one can translate the velocity width of the line to a radius in gravitational radii, $v \approx c(2r_{\rm g}/r)^{1/2}$ \citep{Brown2017}.  Specifically, the velocity at the tidal radius is expected to be $v \approx 43,700 M_6^{1/3} \frac{\rho_\star}{\rho_\odot}^{1/6}$ km s$^{-1}$, where $\rho_\star$ is the average density of the disrupted star \citep{Arcavi2014}.  The unbound tidal debris streams have a lower characteristic velocity, with the most energetic bound material being ejected at speeds of $v \approx 7500 M_6^{1/6}m_\star^{1/3}r_\star^{-1/2}$ km s$^{-1}$ \citep{Strubbe2009}, and should have a narrow profile due to the small velocity dispersion in the tidal tail of unbound debris \citep{Bogdanovic2004}.

PTF-09dj \citep{Arcavi2014} was the first TDE to be fitted by a double-peaked profile to its very broad H$\alpha$ line, albeit the circular Keplerian disk model fit required an extra redshift of 15,000 km s$^{-1}$ to match the wavelengths of the red and blue peaks, and a better fit was achieved, without a bulk redshift, with an elliptical disk, with a large disk inclination, and a pericenter orientation nearly vertical to the observer \citep{Liu2017}.   Furthermore, \cite{Cao2018} argue that even single-peaked broad line profiles in TDEs can be fitted using extremely eccentric disk models with a pericenter position nearly pointing towards the observer and a low disk inclination.

The double-peaked profile of the H$\alpha$ line in PS18kh was modeled by \cite{Holoien2019b} with an elliptical disk with an eccentricity of $e = 0.25$.  However, \cite{Hung2019} favored an expanding spherical outflow model with a flat-topped shape to fit the line profile, because of the identification of narrow absorption features in the broad H$\alpha$ line that were mimicking a double-peaked profile.  The most extreme case for a double-peaked line profile is that observed in the Balmer lines of AT2018hyz/ASASSN-18zj, which are well fitted with a circular disk model in some epochs, but with evolution in the line profile that requires an additional variable Gaussian broad-line component \citep{Short2020, Hung2020}.  The flat Balmer decrement observed in this source (H$\alpha$/H$\beta \sim 
1.5$) was also argued to be the signature of collisional excitation of the line emission, possibly in the chromosphere of the disk.  

One model prediction is that in the presence of an optically thick wind, the lines should be broadened due to electron scattering out of the photosphere \citep{Roth2018}.  Indeed, an observed trend for the broad TDE lines is that they decrease in line-width with time.  \citet{Holoien2016} noted that this is the opposite as expected from reverberation of a fading continuum.  In an AGN, the line-width is determined by the kinematics of the broad-line gas.  Thus, as the continuum fades, you would expect broader lines due to the decreasing radius, and thus higher gas velocities, for which the virialized gas around the black hole was photoionized.  If instead, line width is set by electron scattering, then the evolution of the line width depends on the evolution of the optical depth of the scattering medium.  In the unified TDE model presented by \citet{Dai2018}, due to the wind launched by the super-Eddington accreting debris disk, the optical depth to electron scattering is viewing angle dependent, increasing from the poles to the plane of the disk.  Disentangling these effects is critical for using the line shapes to extract the geometry and kinematics of the photoionized gas.

\subsection{UV Spectroscopy}

UV spectroscopy is a precious resource, since it requires observations from space to evade absorption from the Earth's atmosphere.  The {\it HST}/STIS instrument has played an important role in revealing the spectroscopic properties of TDEs in the UV.  As of now, there are only 4 TDEs with UV spectra (Figure \ref{fig:uv}), ASASSN-14li \citep{Cenko2016}, iPTF-16fnl \citep{Brown2018}, iPTF-15af \citep{Blagorodnova2019}, and AT2018zr/PS18kh \citep{Hung2019}.  They are characterized by a hot, continuum, and strong, broad UV-resonance lines, in emission and sometimes absorption, including Ly$\alpha$, N V $\lambda 1240$, Si IV $\lambda 1400$, and C IV $\lambda 1550$.  They have many spectral features in common with the broad absorption line quasar (BALQSOs), but are notable for the higher strength of nitrogen features relative to carbon, and the relative weakness of C III] $\lambda 1909$, and Mg II $\lambda 2796,2804$, similar to the rare class of "Nitrogen-rich QSOs".

\begin{figure}[t]
\includegraphics[width=4in,trim= 2cm 8cm 0 8cm, clip]{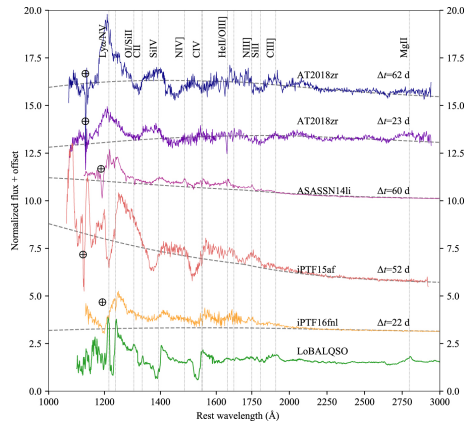}
\caption{\footnotesize {\it HST}/STIS UV spectroscopy of TDEs.  Broad absorption and emission UV-resonance lines are detected on top of a hot, blue continuum, including Ly$\alpha$, N $V \lambda 1240$, Si $IV \lambda 1400$, and C $IV \lambda 1550$.  A UV spectrum of a low-ionization broad absorption line quasar (LoBALQ) is shown for comparison.  Figure from \cite{Hung2019}.}
\label{fig:uv}
\end{figure}

Interestingly, \cite{Yang2017} argue that given that C and N are co-located in the same photoionization zone, their ratio is not as sensitive to photoinization conditions (density fo the gas or shape of the photoionizing continuum), but rather their relative chemical abundances.  The measured abundance ratios measured for nitrogen to carbon from UV lines of C III] $\lambda$ 1908/N III] $\lambda 1750$, indicate N-enriched material.  Indeed, \cite{Kochanek2016} argue that the disrupted star's mass and evolutionary state on the main-sequence will be imprinted on the TDE spectrum, including the enhancement of He, and the rapid enhancement of N and the depletion of C in the CNO cycle for stars more massive than $1 M_\odot$.

Recently, \cite{Parkinson2020} came up with a unification scheme for TDE UV spectra using a Monte Carlo ionization and radiative transfer code, in which the lines are produced in a biconical accretion disk wind, and broad absorption lines are produced for sight lines into the wind cone, and broad emission lines are produced otherwise.  This model successfully reproduces the diversity of UV spectra shown in Figure \ref{fig:uv}, including the suppression of C IV $\lambda 1550$ relative to N V $\lambda 1240$ due to CNO-processed chemical abundances of the stellar debris.

\subsection{Infrared Dust Echoes} \label{sec:mir}

  In the presence of dust, the UV-bright flare of a TDE will be absorbed and re-emitting in the mid-infrared, producing a luminous "echo" \citep{Lu2016}.  Indeed, archival searches of {\it Wide-field Infrared Survey Explorer (WISE)} data revealed luminous, delayed mid-infrared (MIR) flares from three TDEs, including ASASSN-14li \citep{vanVelzen2016, Jiang2016}, PS16dtm \citep{Jiang2017}, and OGLE17aaj \citep{Yang2019}.  In a study of quiescent galaxies with variable MIR emission measured by {\it WISE}, there were 14 candidates for long-term fading that could be attributed to a TDE \citep{Wang2018}.
The detection of increased Fe II emission in several of the TDEs with dust echoes was explained by Fe released into the gas phase from dust grains by UV photons, providing direct evidence for dust sublimation by the TDE flare \citep{Jiang2019}.

One of the advantages of detecting a dust echo, is that the luminosity in the MIR can be used as a bolometer, to determine the total luminosity emitted by the TDE, even without direct access to the unobservable portions of the SED, such as the EUV, where the bulk of the TDE emission is expected to be emitted.  \cite{vanVelzen2016} show that from the estimation of the dust reprocessing shell, $R$, assuming it has the size of the sublimation radius, it will have a radius of

\begin{equation}
R_{\rm dust} \approx 0.15 \left (\frac{L_{45}}{a_{0.1}^2 T_{1850}^{5.8}} \right )^{1/2}~{\rm pc},
\end{equation}

\noindent where $L_{45} = L/10^{45}$ erg s$^{-1}$ is the luminosity of the flare, $a_{0.1} = a/ 0.1 \mu$m, where $a$ is the size of the dust grains, and $T_{1850} = T_{\rm dust}/1850$~K. Thus from measuring the dust temperature from the MIR colors, and inferring the radius from the time delay of the dust echo with respect with the TDE flare, where $\tau \sim R_{\rm sub}/c$ \citep{Jiang2016}, one can infer the peak luminosity over all frequencies where dust absorbs radiation from the TDE.  Then the covering factor of the dust can be determined from the ratio of the total energy radiated by the dust in the MIR by the energy absorbed by the dust, $f_{\rm dust} = E_{\rm dust}/E_{\rm abs}$.  One interesting trend is that $f_{\rm dust}$ appears to be quite low for TDEs, $\sim 1\%$ \citep{vanVelzen2016}.  This is much smaller than the covering fraction inferred from MIR flares from changing-look AGN (CLAGN).  CLAGN are AGN which transform in spectral type, broad-line (Type 1) to narrow-line (Type 2), and vice versa, accompanied by a luminous flare of continuum radiation.  CLAGN are expected to have a pre-existing dusty torus, and thus their characteristic higher values of $f_{\rm dust}$ could be a potential discriminator between TDEs vs. CLAGN for the source of the flaring continuum radiation \citep{Frederick2019}.

\subsection{Coronal-line Emitters}
Another manifestation of a TDE is line emission powered by gas in the vicinity of the black hole that is photoionized by the flare of radiation from the TDE.  The class of "extreme coronal-line emitters" (ECLEs), galaxies with transient, fading narrow-line emission of [Fe VII], and coronal lines from [Fe X] to [Fe XIV], and in some cases broad recombination lines of H$\alpha$, H$\beta$, and He II $\lambda 4686$, have been argued to be the result of photoionization by a TDE \citep{Komossa2008, Wang2012, Yang2013}.  \cite{Dou2016} also found fading mid-infrared emission associated with these coronal-line emitter TDE candidates, further evidence that these were powered by TDEs in a gas-rich environment.  In fact, in \cite{Palaversa2016}, they were able to recover the nuclear flare that powered the high-ionization lines in one of the ECLEs from archival LINEAR optical monitoring data, as well as follow-up {\it Swift} observations of archival \textsl{GALEX} UV detection, which confirmed the fading of the UV emission after the flare.  In general, ECLE's allow us to probe the effects of TDEs on much longer timescales than the accretion event itself, thanks to light travel time delays as the flare propagates through a gas-rich circumnuclear medium \citep{Wang2012}.  Future, massively multiplexed fiber spectrographic surveys could yield a large population of ECLEs, and place a valuable, independent constraint on the TDE rate.

\subsection{X-rays} \label{sec:xray} 

There are now 12 UV/optically-selected TDEs which have been detected in the soft X-rays:  GALEX D3-13, GALEX D1-9, ASASSN-14li, ASASSN-15oi, AT2018zr/PS18kh, AT2018hyz/ASASSN-18zj, AT2019azh/ASASSN-19dj, AT2019dst/ZTF19aapreis, AT2019ehz/Gaia19bpt, AT2018fyk/ASASSN-18ul, and OGLE16aaa.  While the spectral properties of the soft X-ray component in these TDEs share the same characteristics of the soft X-ray selected TDEs: an extremely soft spectrum well fitted by blackbody temperatures that range from $0.02-0.13$ keV (see Figure \ref{fig:tde_year}), the relative brightness of this component compared to the UV/optical component can range from 1 to 10$^{3}$ \citep{Gezari2017, Wevers2019, vanVelzen2019, vanVelzen2020}, and also shows dramatic variability with time, that is unlike the smoothly evolving power-law decline of the UV/optical flare, and is often completely uncorrelated.  Understanding what is driving the wide range and variability of the optical to X-ray ratio in TDEs may be the most revealing diagnostic for what is powering these two components: accretion, reprocessing, or stream-stream collisions.

In the case of ASASSN-15oi \citep{Gezari2017, Holoien2018} and AT2019azh/ASASSN-19dj \citep{Liu2019, vanVelzen2020} there was a gradual brightening in the soft X-rays over a timescale of a year, which the UV/optical component was declining from peak.  In the case of AT2018fyk/ASASSN-18ul, the TDE is detected in the soft X-rays at early times, but with a factor of $2-5$ variability on the timescale of days, and then shows brightening at late times, but then shows a dramatic brightening by a factor of 10 in just 6 days, 3 months after the peak of the flare, resulting in a rapid decrease in the optical to X-ray ratio \citep{Wevers2019}.  OGLE16aaa \citep{Kajava2020} showed the rapid appearance of soft X-rays 6 months after the optical peak, followed by a decline to a lower, steady-state soft X-ray flux over the next year.  

In order to characterize the relative strength of the UV/optical and soft X-ray components, I use commonly used definition of $\alpha_{\rm ox}$ for characterizing the SEDs of AGN, but modify it to measure the ratio of the monochromatic luminosity density at $2500~{\rm \AA}$ to 0.5 keV, instead of 2 keV, since most of the TDEs show such soft spectra, that they are not detected above 1 keV.  I use the unabsorbed X-ray luminosity density corrected for the Galactic HI column density, and the UV luminosity for the Galactic extinction.  Thus I define, 

\begin{equation}
\alpha_{\rm osx} = \frac{\log[L_\nu (2500~{\rm \AA})/L_\nu(0.5~{\rm keV})]}{\log [\nu(2500~{\rm \AA})/\nu(0.5~{\rm keV})]}
= 0.5 \log[L_\nu (0.5~{\rm keV})/L_\nu(2500~{\rm \AA})].
\end{equation}

\noindent In the case where $\nu L_\nu (0.5~{\rm keV}) = \nu L_\nu (2500~{\rm \AA})$, one gets $\alpha_{\rm osx} = -1$.  In AGN, there is an anti-correlation between $\alpha_{\rm ox}$ and $L_\nu (2500~{\rm \AA})$ \citep{Maoz2007}, although in this case $\alpha_{\rm ox}$ is measured at 2 keV, and is comparing the relative strength of UV emission from the "big blue bump" associated with thermal emission from the accretion disk, and non-thermal hard X-ray emission associated with the AGN "corona".  It can be seen in the 5 TDEs with well-sampled UV and soft X-ray light curves from {\it Swift} follow-up observations in Figure \ref{fig:alpha}, that there is strong variability in $\alpha_{\rm osx}$ of the TDEs, with a general trend of increasing $\alpha_{\rm osx}$ with time, suggestively peaking near $\alpha_{\rm osx} = -1$.  In the case of AT2019ehz/Gaia19bpt, the soft X-ray flux appears to flare by a factor $> 10$ on timescales of just a few days, with $\alpha_{\rm osx} \sim -1$ during each flare \citep{vanVelzen2020}.  In the case of ASASSN-15oi and AT2019azh/ASASSN-19dj, the soft X-ray flux increases by a factor $> 10$ over 200 days, with $\alpha_{\rm osx}$ increasing from $-2.2$ to $-1.0$, but with no evidence of a change in absorbing column density, and no significant evolution in temperature \citep{Gezari2017, Hinkle2020b}.  
AT2018fyk/ASASSN-18ul showed variability of a factor of $2-5$ on the timescale of days, and then, 100 days after peak, brightens by a factor of 10 in less than a week \citep{Wevers2019}, approaching close to $\alpha_{\rm osx} \sim -1$, but again with no evidence of a change in line-of-sight absorption. 

Understanding what is driving this variability on both short ($\sim$ day) and long ($\sim$ yr) timescales is critical for determining if variable obscuration and/or delayed accretion is determining the optical to soft X-ray flux ratio.  Already, the lack of an obvious decrease in line-of-sight column density with increasing soft X-ray luminosity is problematic for the obscuration scenario.  However, it may be that if the opacity is dominated by electron scattering by a highly ionized medium \citep{Dai2018}, then the soft X-rays can be obscured without changing the observed spectral slope as is the case for absorption by neutral gas.  

\begin{figure}[t]
\includegraphics[width=3in,trim=1cm 0 4cm 14cm, clip]{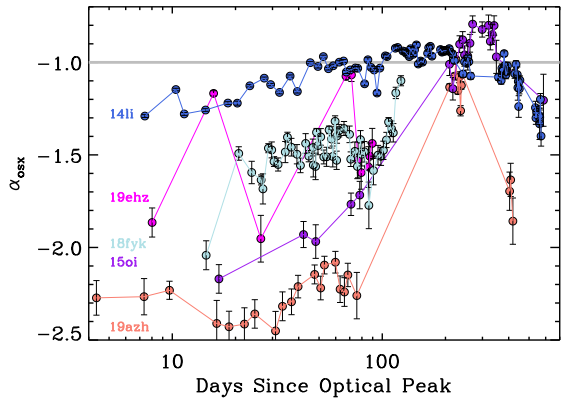}
\caption{\footnotesize Dramatic X-ray variability of TDEs detected with both a UV/optical and soft X-ray component, characterized with the parameter $\alpha_{\rm osx}$, and measured from {\it Swift} XRT and UVOT/uvw2 observations of ASASSN-14li, ASASSN-15oi, AT2018fyk/ASASSN-18ul, AT2019azh/ASASSN-19dj, and AT2019ehz/Gaia19bt.  Grey line shows the characteristic value of $\alpha_{\rm osx} = -1$ which corresponds to $ \nu L_\nu (2500~{\rm \AA})/\nu L_\nu (0.5~{\rm keV}) = 1$.} 
\label{fig:alpha}
\end{figure}

\subsection{Evidence for Jets and Outflows} \label{sec:outflow}

One of the biggest surprises in TDE observations was the discovery of a class of ``relativistic" TDEs, whose luminous, non-thermal radiation was explained as viewing a jet face on that was launched by a TDE.  In retrospect, this class of TDE is a natural extension of the ``blazar'' class for AGN, and the impact of jets on the radiation from TDEs had already been discussed in the literature in the context of jets interacting with the circumnuclear medium.  The detection of such radio emission was not actualized until the low-redshift optically selected TDE ASASSN-14li.  For the first time, a multiwavelength campaign was able to detect several components of emission from this TDE across the electromagnetic spectrum, including transient radio emission.

The radio emission from ASASSN-14li was interpreted as synchrotron emission from either {\it external shocks} driven by the interaction of a non-relativistic outflow \citep{Alexander2016}, the unbound debris streams \citep{Krolik2016}, or a relativistic jet \citep{vanVelzen2016} with the circumnuclear medium, or from {\it internal shocks} in a relativistic jet \citep{Pasham2018}.  However, the detection of a correlation between the X-ray and radio light curves, with $L_{\rm radio} \propto L_{\rm X}^2$, suggests that the accretion and jet power are linearly coupled \citep{Pasham2018}.  This follows from the fact that the synchrotron luminosity from a power-law distribution of electrons with index $p$ in the radio scales with the jet power ($Q_j$) as $L_{\rm radio} \propto Q_j^{1+\frac{p+1}{4}}$.  Thus for $p=3$ and $L_{\rm radio} \propto L_{\rm X}^2$, where $L_{\rm X} \propto \dot{M}_{\rm acc}$ one gets $Q_j \propto \dot{M}_{\rm acc}$.  \cite{Pasham2018} also observe a time lag between soft X-ray and the radio of 12 days, which is self-consistent with their model for a freely expanding canonical jet.  The strong coupling between the accretion rate and jet power strongly favor the internal jet model over an external emission mechanism \citep{Pasham2018, Bright2018}.  

Radio emission has also been detected from XMMSL1 J0740-85\citep{Alexander2017} and AT2019dsg \citep{Stein2020}, suggesting that radio follow-up observations, or even radio blind-searches for transient radio emission \citep{Anderson2019}, can place important constraints on the nature and frequency of jets and outflows in TDEs.  Already one can infer from radio follow-up that high radio-luminosity, relativistic jets similar to Sw J1644+57 must occur in only 1\% of TDEs \citep{Alexander2020}.   

The super-Eddington accretion rates in TDEs make them especially conducive to launching an outflow.  This outflow could manifest itself as bright optical emission from the reprocessing of higher energy emission from the accretion disk \citep{Strubbe2009, Lodato2011, Metzger2016}, or in the form of absorption lines, particularly in the UV \cite{Strubbe2011}.  Highly ionized outflows have now been detected from spectra in the X-rays from ASASSN-14li \citep{Miller2015, Kara2018}, and in the UV from ASASSN-14li \citep{Cenko2016}, iPTF16fnl \citep{Brown2018}, iPTF15af \citep{Blagorodnova2019}, and AT2018zr/PS18kh \citep{Hung2019}, with a wide range of velocities, sometimes blue-shifted and sometimes red-shifted, reflecting a diverse set of physical conditions in the outflow in terms of their geometry, kinematics, and optical depth \citep{Hung2019}. Progress in mapping the UV spectra of TDEs to the geometry and orientation of the accretion disk wind is critical for testing models in which the optical-to-X-ray ratio is determined by the inclination angle to a radiatively-driven wind \citep{Dai2018}.   In general, understanding the radiation, winds, and jets from TDE accretion disks give important insight to the feedback mechanisms of AGN in general.  

\section{TENSIONS BETWEEN OBSERVATIONS AND TDE THEORY}
\subsection{Large Inferred Radii}

One of the common properties of both X-ray and UV/optically selected TDEs is the thermal nature of their continuum.  This is expected for the continuum from the circularized, accreting debris disk.  Following the classical relations for an optically, thick, geometrically, thin accretion disk. one expects a peak effective temperature of \citep{Miller2015}:

\begin{equation}
T_{\rm eff, peak} = 0.54 \left ( \frac{\dot{M}}{\dot{M_{\rm Edd}}} \frac{G^2 M_{\rm BH}^2}{\kappa \sigma \eta c R_{\rm in}^3} \right ) ^{1/4} \approx 4 \times 10^{5} K \left ( \frac{\dot{M}}{\dot{M_{\rm Edd}}} \right ) (\eta/0.1)^{-1/4} M_{\rm BH}^{-1/4}
\end{equation}

\noindent where $\kappa$ is the opacity due to electron scattering.  Both components are well described by a blackbody, albeit with a factor of $\sim 10$ difference in temperature.  The first TDE candidates, discovered in the soft X-rays by the ROSAT survey, initially were puzzling due to their relatively small inferred radii, using the simple relation for a blackbody, $L = 4\pi\sigma R_{\rm bb} T^4 $, even smaller than the Schwarzschild radius of the central black hole.  However, the next generation of TDE candidates, discovered in UV and optical surveys, had the opposite problem.  Their inferred radii were too large compared to the expectations of the size of the circularized debris disk, with $R_{\rm disk} = 2R_{\rm T}$.  These disagreements with the basic theoretical predictions of accretion powered emission from a circularized debris disk has necessitated the development of more complex theories, involving a reprocessing envelope \citep{Loeb1997, Guillochon2014}, electron scattering \citep{Li2002}, stream-stream collisions \citep{Piran2015, Jiang2016}, and radiatively driven winds \citep{Metzger2016}.  

In Figure \ref{fig:rad} I show the radius of the observed blackbody emission in the 49/56 of the TDEs in Table \ref{table1} with a black hole mass inferred from the host galaxy mass measured from the broadband SED for 10 of the X-ray TDE hosts \citep{Graur2018, Wevers2019, French2020} and 39 of the UV/optical TDE hosts \citep{vanVelzen2018, vanVelzen2020}, to estimate $M_{\rm BH}$.  I use the correlation between the total galaxy mass  ($M_{\rm gal}$) and $M_{\rm BH}$ measured from dynamical masses of early and late-type galaxies from \cite{Greene2019}, $\log(M_{\rm BH}/M_\odot) = 7.56 + 1.39[\log(M_{\rm gal}/M_\odot) - 10.48]$,
with an intrinsic scatter of 0.79 dex.  This relation is most relevant for estimating $M_{\rm BH}$ for the TDE hosts since it is based on the total galaxy mass, and thus does not require knowledge of the bulge-to-total ratio, which is not well measured for most of the host galaxies.  However, the relation is based on only 5 galaxies below $10^{10} M_\odot$ with dynamical BH masses, and there appears to be two different normalizations between the early type and late-type galaxies.   The $M_{\rm BH}$ versus $\sigma_\star$ relation has a smaller scatter (0.53 dex) and does not have this difference in normalization between the galaxy types.  Thus, continued high spectral resolution follow-up of TDE galaxy hosts is critical for the best constraints on $M_{\rm BH}$. 

On this plot are also the expectations for the various characteristic size scales for a TDE as a function of $M_{\rm BH}$, including, $R_{\rm S} = 2GM_{\rm BH}/c^2$, and

\begin{equation}
R _{\rm circ} = (1 + e_{0}) R_{\rm p} \sim 2R_{\rm T}
\end{equation}

\noindent circularization radius due to angular momentum conservation, where $e_{0}$ is the initial eccentricity and $R_{\rm p}$ is the pericenter of the star's orbit.

\begin{equation}
R_{\rm I} = \frac{a_{\rm mb} (1-e_{\rm mb}^2)}{(1-e_{\rm mb} \cos({\phi/2}))} 
\end{equation}
 
\noindent self-intersection radius from \citet{Dai2015}, where $e_{\rm mb} \approx 1 - (2/\beta) \left ( \frac{M_\star}{M_{\rm BH}} \right) ^{1/3}$ is the eccentricity of the orbit of the most bound debris, and $\phi = \frac{6\pi G M_{\rm BH}}{a_{\rm mb}(1-e_{\rm mb}^2)c^2}$ is the angle of apsidal precession, and $a_{\rm mb} = R_{\rm T}^2/(2R_\star)$ is the semi-major axis of the most bound debris \citep{Dai2015, Bonnerot2016}.

\begin{figure}[t]
\includegraphics[width=3in,trim=1cm 0 4cm 14cm, clip]{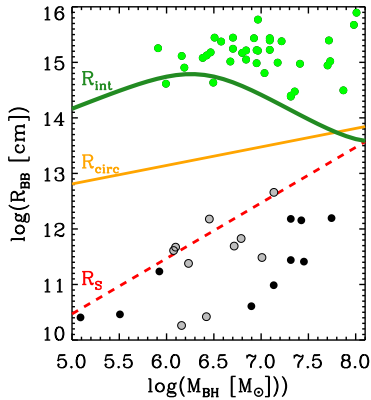}
\caption{ \footnotesize Blackbody radius as a function of black hole mass estimated from the total galaxy mass for UV/optically selected TDEs (green), X-ray selected TDEs (black), and the soft X-ray component of UV/optically selected TDEs (grey).  While the radius of the soft X-ray component appears to track the increase of the Schwarzschild radius linearly, with mass, $R_{\rm S} \propto M_{\rm BH}$ (plotted with a red dashed line) the UV/optical component appears not to have a strong dependence on black hole mass, is a factor of 100 larger than the expected size of the circularized debris disk, $R_{\rm circ} \sim 2R_{\rm T}$ (plotted in yellow), and is more consistent with the self-intersection radius of the debris streams (plotted in green for an impact parameter $\beta = 1$), and which also has a weak dependence on $M_{\rm BH}$.}
\label{fig:rad}
\end{figure}

It is clear that the characteristic radii of the UV/optical thermal component in TDEs are much larger than expected for a  debris disk formed from the circularized debris streams, and that the soft X-ray thermal component in TDEs is often even smaller than the size of the event horizon of the black hole ($R_{\rm S}$).  Both of these situations are problematic, and require an expansion of the simplest TDE models.  Interestingly, the UV/optical component is close to the size of the self-intersection radius of the debris streams, as pointed out by \citet{Wevers2019}.  An important consideration is also whether or not the UV/optical component is powered by reprocessing of the soft X-ray component, or is in fact a source of emission generated from the shock interaction of the debris streams themselves. In the following section, we examine the relative behavior of these two components in order to determine their origin.

\subsection{Large Optical to X-ray Ratio, and its Evolution With Time} \label{sec:opt_xray}

One of the most intriguing questions in TDE observations is why are many optically selected TDEs faint in the X-rays, and what is driving the X-ray variability observed in several optically selected TDEs.  The main explanation for the high optical-to-X-ray ratio in TDEs has been either a) obscuration or b) delayed accretion.  In the case of an optically thick wind, the material is initially opaque to X-rays due to electron scattering.  It is not until the ejecta expands, and becomes completely ionized by the inner disk radiation after $t_{\rm ion} \approx 2 t_{\rm fb} \beta^{-3.4} M_6^{-1.1} m_\star^{0.71}$ that the soft X-rays can escape \citep{Metzger2016}, implying that for the largest black holes, $t_{\rm ion} << t_{\rm fb}$, and the X-ray radiation escapes promptly, resulting in an "X-ray luminous" TDE.    
However, viewing angle effects may come into play if the geometry of the accretion flow allows for some lines of sight out of the wind and directly towards the inner accretion disk \citep{Dai2018}.

Another nice outcome of this model, is that the radiation diffused through the wind results in an effective temperature of,

\begin{equation}
T_{\rm eff} = \left ( \frac{L_{\rm rad}}{4\pi \sigma R_{\rm wind}^2} \right )^{1/4} \approx 2.1 \times 10^{4} {\rm K} (\eta/0.1)^{1/4} \beta^{-1/4} m_\star^{13/60} M_6^{-13/24} \left (\frac{t}{t_{\rm fb}} \right )^{-11/12}.
\end{equation}

\noindent and provides a potential explanation for the observed temperature of the UV/optical component in TDEs, but not their relatively constant temperature with time, with several TDEs showing an {\it increase} in temperature with time \citep{vanVelzen2020, Hinkle2020}, challenging these models even further.  However, this model's prediction for a cooling of the temperature with time does not take into account the fact that as the material expands, and becomes optically thin, the photosphere will recede to deeper layers, resulting in a slower evolution of $T_{\rm eff}$. 

However, there are more complications.  This model assumes that the material circularizes and forms an accretion disk with $t_{\rm circ} << t_{\rm fb}$.  In fact this process depends sensitively on the central black hole mass.  
In several numerical and analytical studies, most recently by \citet{Bonnerot2020}, it is the smaller mass black holes for which the circularization timescale is quite long \citep{Bonnerot2016}, 

\begin{equation}
t_{\rm circ} = 8.3 t_{\rm fb} M_6^{-5/3} \beta^{-3} 
\end{equation}

\noindent and this was dubbed a "Dark Year for Tidal Disruptions" by \citet{Guillochon2015}.  The smaller black holes also result in stream-stream collisions further from the black hole, resulting in longer viscous timescales, since the debris that accretes onto the black hole from that self-intersection point is subject to the viscous timescale at that distance:

\begin{equation}
t_{\rm visc} = \alpha^{-1} (h/r)^{-2} P_{\rm circ}
\end{equation}

\noindent where $P_{\rm circ} = 2\pi \sqrt{a_{\rm circ}^3/GM_{\rm BH}}$.  If $t_{\rm fb} << t_{\rm visc}$ then the accretion rate will be "slowed" with respect to the fallback rate, also resulting in a reduced peak luminosity.

An alternative model to explaining the weak X-ray luminosity at early times in optically-selected TDEs, is that the optical emission is powered locally by shock heating from stream-stream collisions near apocenter, which thermalizes, and diffuses out through the infalling matter with an intrinsic temperature of 

\begin{equation}
T \sim 3.3 \times 10^{4} m_\star^{1/10}M_6^{-3/8}~{\rm K},
\end{equation}

\noindent thereby matching the observed temperature of the UV/optical component in TDEs, without requiring the reprocessing of soft X-radiation from an accretion flow.  In this model the UV/optical emission and the soft X-ray emission are completely decoupled, and one could explain the late-time brightening of the soft X-ray emission in some TDEs as delayed accretion in a circularizing debris disk.  There already is some evidence for a black hole mass dependence on the emergence of a late-time UV plateau in optically selected TDEs, explained as delayed emission from a viscously spreading accretion disk in the TDEs with lower mass black holes ($M_{\rm BH} < 10^{6.5} M_\odot$) \citep{vanVelzen2019}.  This result important in that it also implies that the prompt UV/optical component of these TDEs is not powered by accretion!

\subsection{Correlations (or Lack Thereof) with Central Black Hole Mass} \label{sec:bh}

One of the robust predictions of TDE theory, is that there should be a maximum central black hole mass, above which a star will cross the event horizon before being disrupted, i.e., $R_{\rm T} > R_{\rm S}$, which occurs for

\begin{equation}
 M_{\rm BH} < \frac{c^3}{M_\star^{1/2}}\left( \frac{R_\star}{2G} \right )^{3/2} = 1.1 \times 10^{8} M_\odot m_\star^{-1/2} r_\star^{3/2}\
 \end{equation}

\noindent where $r_\star = R_\star/R_\odot$ and $m_\star = M_\star/M_\odot$.  \citet{vanVelzen2018} showed that indeed the luminosity function of TDEs as a function of black hole mass (measured from the host galaxy bulge velocity dispersion) did  show an intrinsic cut-off at $\sim 10^{8} M_\odot$, and argued this as clear evidence for the black hole event horizon.  The one exception to this cut-off was ASASSN-15lh, a TDE candidate (also argued to be a SLSN by \cite{Dong2016}) from the center of a passive red galaxy with an inferred central black hole mass (from the host galaxy stellar mass) of $\sim 6 \times 10^{8} M_\odot$, well above the Hills mass of a non-rotating (Schwarzschild) black hole for a solar-type star.  However, a star could be disrupted outside the event horizon for this black hole mass if either it was a maximally spinning (Kerr) black hole, or if the star was massive ($M_\star > M_\odot$).  However, given the old stellar population of the host galaxy measured from stellar population synthesis fits to its spectral energy distribution, a spinning black hole was favored \citep{Leloudas2016}.  

Furthermore, given that the tidal disruption radius, fallback timescale, and peak mass accretion rate have dependencies on black hole mass, one would expect the properties of the TDE flares to correlate with $M_{\rm BH}$.  Currently, the most stringent constraints on $M_{\rm BH}$ for TDE hosts come from stellar velocity dispersion measurements.  \citet{Wevers2019} reported black hole mass estimates for 23 TDE hosts using homogeneous measurements of the flux-weighted central stellar velocity dispersion ($\sigma_\star$) from medium resolution ($\Delta v \sim 15 - 60$ km s$^{-1}$), optical long-slit spectra, and found a relatively flat distribution of $M_{\rm BH}$ between $5 < \log(M_{\rm BH}/M_\odot) < 8$, with no significant difference in black hole masses between the optically selected and X-ray selected TDEs.

However, the ultimate goal for TDE observations is to use the properties of the TDE itself to weigh the mass of the central black hole, instead of relying on indirect host galaxy scaling relations.  Indeed, \cite{Mockler2019} attempted to do so by fitting multi-band optical light curves with a model based on numerical simulations of the fallback rate ($\dot{M}_{\rm fb}$) as a function of stellar polytropic index ($\gamma$) and impact parameter ($\beta$), and fitting for an accretion efficiency ($\eta$), where $\dot{M}_{\rm fb} = L/(\eta c^2)$, and adjusting $\eta$ such that the luminosity cannot exceed the Eddington limit.   In order to fit the  temperatures and constant colors characteristic of optically-selected TDEs, the model includes the addition of a reprocessing photosphere with a radius that can change with a power-law dependence on the luminosity, and with a minimum and maximum radius set by the innermost stable circular orbit of the black hole ($R_{\rm isco}$) and the semi-major axis of the debris streams ($a_0$), respectively.  Finally, the model also allows for a ``viscous delay", to account for potential delays between the fallback of material, and its subsequent accretion via an accretion disk.  This parametric model, fitted with an MCMC routine to find the highest likelihood model parameters with Bayesian statistics (\texttt{MOSFiT}) yields a constraint on $M_{\rm BH}$ and $M_\star$ \citep{Mockler2019}.  However, the black hole masses derived from \texttt{MOSFiT} show just a marginal agreement with the black hole masses inferred from the host galaxies' properties.  This is not surprising, since it is not clear yet that the observable parameters of a TDE, such as rise time or peak luminosity, even scale with $M_{\rm BH}$ as expected for a light curve powered solely by fallback accretion.

\begin{figure}[t]
\begin{subfigure}[t]{0.45\textwidth}
\centering
\includegraphics[width=\textwidth,trim=0cm 0 0.cm 0]{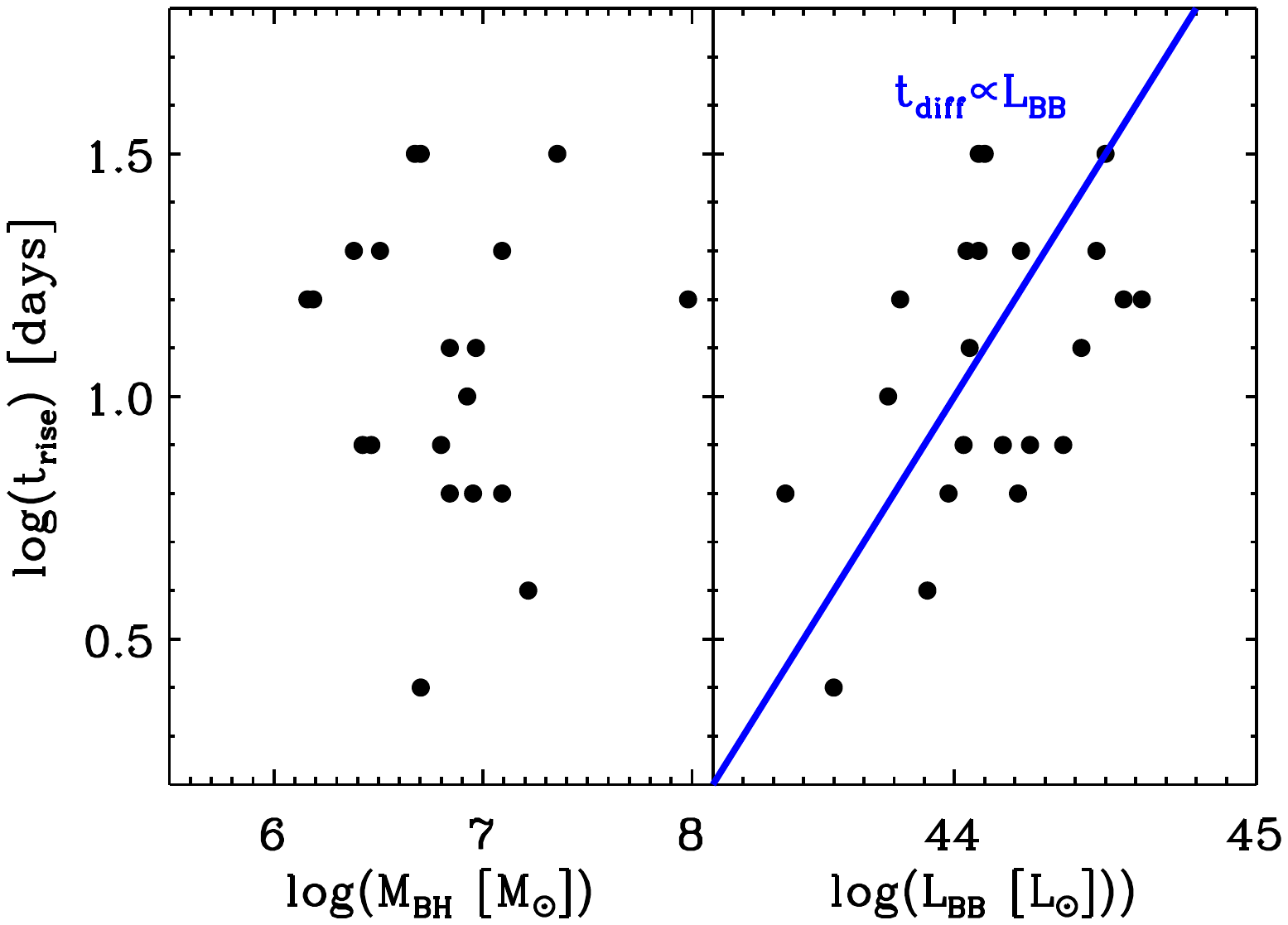}
\end{subfigure}
\begin{subfigure}[t]{0.45\textwidth}
\centering
\includegraphics[width=\textwidth,trim=0.cm 0cm 0.cm 0, clip]{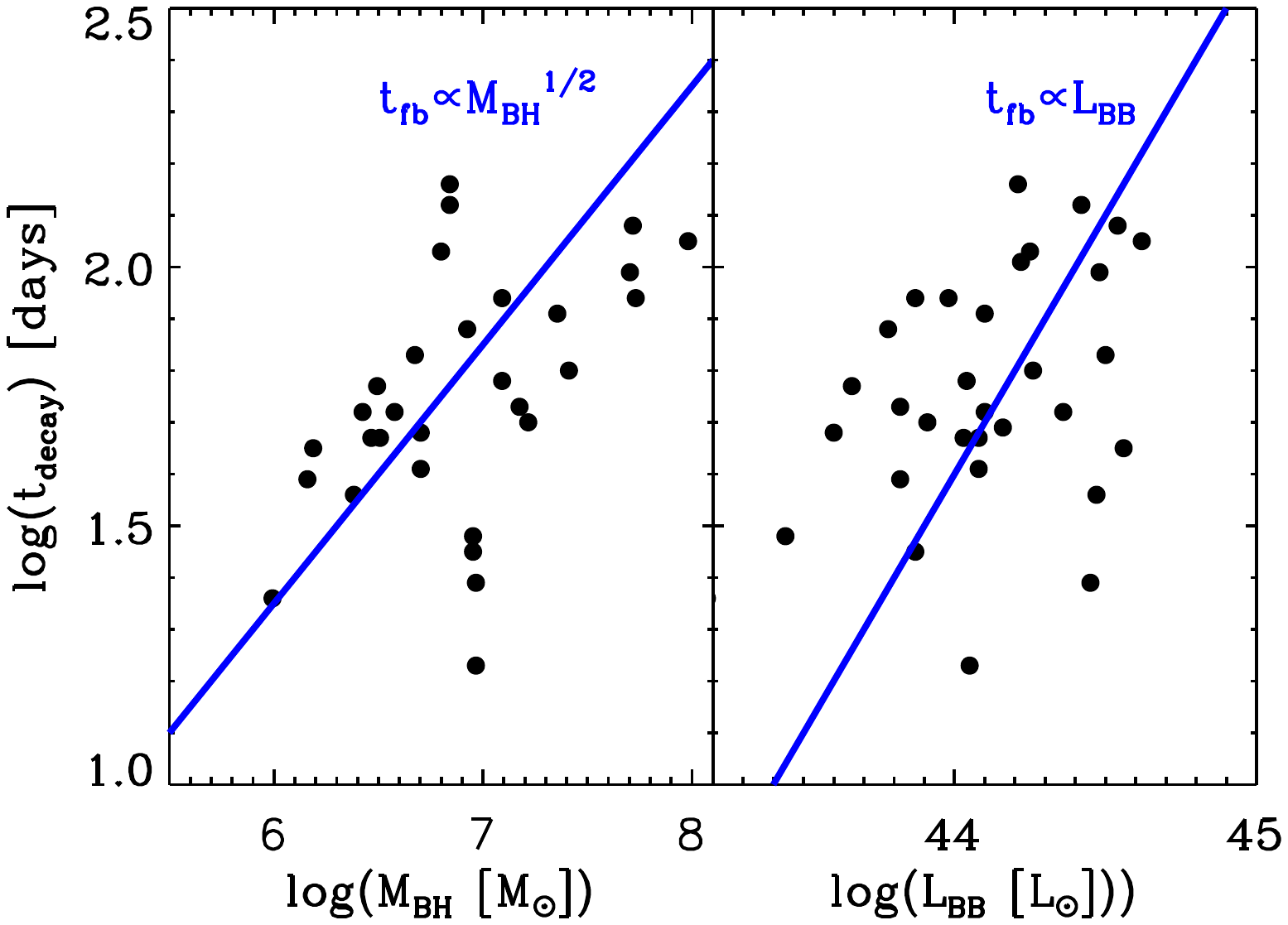}
\end{subfigure}
\caption{\footnotesize {\it Left:}  Correlation (or lack thereof) of rise timescale with black hole mass and peak luminosity for TDEs with black hole mass estimates from their total galaxy mass, and a well-sampled pre-peak light curve.  Correlations for the fallback rate shown in blue. {\it Right:} Correlation of decay timescale with black hole mass and peak luminosity for TDEs with black hole mass estimates from their total galaxy mass.  Correlations for the fallback rate shown in blue.}
\label{fig:rise}
\end{figure}

We now have 22 TDEs with pre-peak sampling of their light curves, PS1-10jh \citep{Gezari2012}, PS1-11af \citep{Chornock2014}, PTF09ge \citep{Arcavi2014}, iPTF16fnl \citep{Blagorodnova2017}, 15/17 TDEs in ZTF \citep{vanVelzen2020}, plus AT2018dyb/ASASSN-18pg \citep{Leloudas2019}, AT2018fyk/ASASSN-18ul \citep{Wevers2019}, AT2019ahk/ASASSN-19bt \citep{Holoien2019} (see Figure \ref{fig:nicholl} for some examples).  This enables us to investigate how their rise time to peak correlates with other TDE parameters and host galaxy properties.  In Figure \ref{fig:rise} I plot the e-folding rise timescale to the peak luminosity, $\Delta t_{e,rise}$, and e-folding decay timescale, $\Delta t_{e,decay}$ as measured by \cite{vanVelzen2020b}, as a function of inferred black hole mass from the host galaxy mass.  As found in \cite{vanVelzen2020}, there is a correlation between luminosity and rise time, but surprisingly no correlation between rise time and inferred central black hole mass.  Interestingly, if the rise time was tracing the fallback time scale, then since $t_{\rm fb} \propto M_{\rm BH}^{1/2}$ and $L_{\rm peak} \propto M_{\rm BH}^{-1/2}$, then one would expect $t_{\rm peak} \propto L_{\rm peak}^{-1}$.  If instead, the rise time is tracing the radiative diffusion timescale, $t_{\rm diff} \propto \rho R^2$, which for a spherical distribution of radius $R$ and mass $M$, translates to $t_{\rm diff} \propto M/R$, then if the radius of the photosphere is set by the intersection radius of the debris streams, $R_{\rm int}$, and the peak luminosity traces the mass accreted, then one gets $t_{\rm diff} \propto L/R_{\rm int}$, and then one would get the positive correlation between $L$ and $t_{\rm rise}$ that is observed. With a similar argument, \cite{vanVelzen2020} explained that the lack of the expected correlation of $t_{\rm peak} \propto M_{\rm BH}^{-1/2}$ could be attributed to photon trapping in the presence of an outflow, which will prolong the rise time to peak if $t_{\rm tr} > t_{\rm fb}$, where $t_{\rm tr}/t_{\rm fb} \propto M_{\rm BH}^{-1/2}$ \citep{Metzger2016}, and thus should be a more important effect for less massive black holes.

It has been proposed, that instead of using fits to the TDE light curve, one can simply use the observed temperature and UV/optical luminosity at peak to infer $M_{\rm BH}$ and $M_\star$, if the UV/optical emission is powered by the dissipation of energy by shocks intersecting near the apocenter of the highly eccentric orbits of the stellar debris, where 

\begin{equation}
a_0 = \frac{G M_{\rm BH}}{\Delta \epsilon} = M_{\rm BH}^{2/3} R_{\star} M_{\star}^{-2/3} \propto M_{\rm BH}^{2/3}M_{\star}^{2/9},
\end{equation}

\noindent and assuming a mass-radius relation of $R_\star \propto M_\star^{8/9}$ \citep{Ryu2020b}.  Since the luminosity from the shock in this model scales as $L_{\rm sh} \sim \frac{G M_{\rm BH} \dot{M}}{a_0} \propto M_{\rm BH}^{-1/6} M_\star^{4/9}$, and the temperature scales as $T \propto [\frac{L_{\rm sh}}{\sigma \Delta \Omega a_0^2}]^{1/4}$, then one gets $T \propto M_{\rm BH}^{-3/8}$ \citep{Ryu2020b}.  Interestingly, this method does provide black hole mass estimates that are in good agreement with the black hole mass estimated from the host galaxy bulge mass.  Another piece of evidence in favor of the physical model where stream-stream shocks at $r \sim a_0$ are powering the UV/optical component, instead of reprocessing of emission from a more compact, circularized accretion disk with $r \sim R_T$.

Another intriguing signal to probe $M_{\rm BH}$, which originates from the heart of the accreting black hole in a TDE, is quasi periodic emission from the innermost stable circular orbit ($R_{\rm isco}$) around the black hole.  Quasi-periodic oscillations have been reported in two TDEs, in an analysis of the non-thermal hard X-ray light curve of jetted TDE Swift J1644+57 \citep{Reis2012} ($P$ = 200 sec), and of the soft thermal X-ray light curve of optically-selected TDE ASASSN-14li  \citep{Pasham2019} ($P=131$ sec).  This period corresponds to a radius of $R = (\frac{GM_{\rm BH} P^2}{4\pi^2})^{1/3} \sim 0.1 (\frac{P}{[100~{\rm sec}]})^{2/3} M_6 R_{\rm S}$, and with a knowledge of the spin parameter $a$, and assuming the emission originates from $R_{\rm ISCO}$, since $R_{\rm isco} = 3R_{\rm S}$ for a=0 to $R_{\rm isco} = R_{\rm S}/2$ for a maximally spinning (a=0.998) black hole, it can be used to determine $M_{\rm BH}$.  

\subsection{Observed Rates:  Too Low, Too High, or Just Right?}

The expected rate of TDEs in galaxy nuclei has been both analytically and numerically determined from calculations of stellar orbits for galaxies of various stellar density profiles (core vs. cusp) as well as black holes of different configurations (Schwarzschild, Kerr, binary, recoiling).  One of the most exciting applications of {\it measuring} the TDE rate as a function of galaxy type and black hole mass, is distinguishing between these scenarios.  For example, measuring the TDE rate on the low-mass end can shed light on the black hole occupation fraction in low-mass galaxies, a potential discriminator between BH seed formation models \citep{Stone2016b}.  Measuring the TDE rate on the high-mass end, can probe the spin distribution of MBHs \citep{Kesden2012b}.  However, while the TDE event rate is well determined, predictions for the {\it observed} tidal disruption flare rate relies on several assumptions:  the peak luminosity, effective temperature, and galaxy central black hole mass.  One can take an empirical approach, and start with the observed luminosities, temperatures, and light curves.  However, there may be a population of TDEs that the current survey selection techniques are missing, or an error in our assumptions about the black hole mass function which is poorly constrained in low-mass galaxies.  Comparing the theoretical rate to the observational rate can thus be quite revealing.  

Measuring the TDE rate ($\dot{N}$) per galaxy requires an understanding of each survey's detection efficiency, $\epsilon \equiv N^{-1} \sum_i^N \epsilon_i$, where $\epsilon_i$ is the detection efficiency for each TDE detection, effective survey time ($\tau$), and number of galaxies monitored ($N_{\rm gal}$), where
$\dot{N} = (N_{\rm TDE}/N_{\rm gal}) \tau \epsilon$ \citep{vanVelzen2014}.  In order to calculate $\epsilon$, one has to make assumptions about the luminosity and time evolution of TDEs.  The first attempts to measure $\dot{N}$ from X-ray and optical survey data resulted in rates of $\sim 10^{-5}$ galaxy$^{-1}$ yr$^{-1}$ \citep{Donley2002, Esquej2008, Gezari2009, vanVelzen2014, Holoien2016}, an order of magnitude below the expectations from dynamical models, but were based on the detection of only $2-3$ TDEs in each survey.  

This tension with the theoretical expectation of the rate from loss-cone dynamics using realistic stellar density profiles of $10^{-4}$ galaxy$^{-1}$ yr$^{-1}$ \citep{Magorrian1999, Wang2004} was initially attributed to selection effects due to insensitivity to faint and fast, or long-timescale TDEs \citep{Kochanek2016}, conservative candidate filtering for SNe or AGN, our uncertainty in the nature of the optical emission component, which may not be present in all TDEs, or an anisotropic velocity distribution of stellar orbits, with a bias towards tangential velocities that would suppress the TDE rate from two-body relaxation \citep{Stone2016b}.

However, with larger samples of TDEs now available, the observed TDE rates appear to be coming into agreement with theoretical expectations.  If one is interested in the volumetric rate ($\dot{n}$), one needs to take into account not the number of galaxies, but the total volume surveyed.  Using the "$1/{\bf V}_{\rm max}$" method, where
${\bf V}_{\rm max} \equiv V(z_{\rm max}) A \times \tau$, and A is the area of the survey, $z_{\rm max}$ is the maximum redshift that one could detect the TDE of a given luminosity by the flux limit of the survey, and ${\bf V}(z_{\rm max})$ is the corresponding volume per unit area.  Using the value of $\sum 1/V_{\rm max}$ in bins of luminosity, \cite{vanVelzen2018} derived a luminosity function of TDEs from 17 TDEs from the literature, well fitted with a steep power-law of the form, 

\begin{equation}
\frac{d\dot{N}}{d \log_{10} L} = \dot{N}_0 (L/L_0)^{a},
\end{equation}

\noindent where $L_{0} = 10^{43}$ erg s$^{-1}$, $\dot{N}_0 \sim 2 \times 10^{-7}$ Mpc$^{-3}$ yr$^{-1}$ and $a \sim -1.5$.
One can then make a more accurate estimate of the TDE rate per galaxy, by calculating $\epsilon$ using the observed luminosity function, instead of a constant characteristic luminosity, or a model-dependent luminosity.  Indeed, \cite{vanVelzen2018} find a per galaxy rate of $1 \times 10^{-4}$ galaxy$^{-1}$ yr$^{-1}$ by using the observed host galaxy stellar mass function from the $1/V_{\rm max}$ method, now coming into agreement with theoretical predictions.  Similarly, \citep{Hung2017} derived a TDE rate from the systematic follow-up of iPTF nuclear transients in the UV with {\it Swift}, resulting in two TDE detections, and assuming the steep luminosity function from \cite{vanVelzen2018} dominated by less luminous flares, resulting in a rate of $\sim 1.7 \pm 10^{-4}$ galaxy$^{-1}$ yr$^{-1}$.  A recent update to the \cite{vanVelzen2018} calculation by \cite{vanVelzen2020b}, with double the sample of TDEs from adding the most recent influx of optical TDE discoveries in the literature, yields a slightly lower overall rate of $6 \times 10^{-5}$ galaxy$^{-1}$ yr$^{-1}$.

One of the challenges for the next generation of surveys, is how to filter transients down to TDE candidates without making assumptions about their properties that would bias the survey against finding the full population of TDEs.  However, despite these challenges, we are finding that the observed TDE rate is in general agreement with dynamical predictions, and with important preferences in the host galaxy types.

\subsection{Do TDEs Prefer Post-Starburst Galaxy Hosts?  If so, Why?}

Given that the TDE rate should be primarily driven by stellar dynamics within the sphere of influence of the central black hole, the rates and properties of TDEs should correlate with host galaxy stellar density and structure.  It was surprising, however, when \cite{Arcavi2014} noted that several TDEs were hosted in a rare subclass of galaxies with no active star formation and a bright population of A stars resulting in strong H$\delta$ absorption known as "E+A" galaxies.  This was broadened further by \citep{French2016} to be a preference for quiescent, Balmer strong galaxies, more common than "E+A" galaxies, but still only making up only $\sim 2$\% of the local galaxy population, but 75\% of the optical TDE hosts known at the time.  This puzzling over-representation of TDEs in this rare galaxy type was attributed to several effects, including a recent galaxy merger (potentially yielding a binary SMBH with an enhanced TDE rate,) or a higher concentration of A stars susceptible for disruption.  

However, TDE hosts also share other properties, including in colors predominantly in the ``green valley" of the mass-color diagram \citep{Law-Smith2017, vanVelzen2020, Hammerstein2020} and high central stellar density \citep{Law-Smith2017, Graur2018, Hammerstein2020}, both favorable conditions for stellar disruptions. 
Using forward modeling techniques, \cite{Roth2020} find that they can explain the lack of ZTF TDE hosts in the blue locus of galaxies in the mass-color diagram from dust obscuration in star-forming galaxies.  However, selection effects alone can not explain their over-representation of E+A galaxy hosts.  Although, \cite{Hammerstein2020} note that the over-representation of the quiescent, Balmer strong, "E+A" (H$\delta > 4$\AA) galaxies dissappears to $\mathcal{O}(1)$ when selecting from only the green and centrally concentrated galaxy population.   

\section{FUTURE PROSPECTS}
\subsection{Improving Search Strategies from the Ground and in Space}

While survey capabilities in the optical band have been steadily increasing, and will have a large jump in survey power with the beginning of the Vera Rubin Observatory Legacy Survey of Space and Time (LSST) in just a couple years, our selection strategies for discovering TDEs will still be hampered by intensive filtering of more common interlopers, such as AGN and SNe.  One feature of TDEs that has been critical for their confirmation from optical candidates has been their uniform property of bright, persistent UV emission.  In Figure \ref{fig:uvtde} I show the UV-optical color vs. their absolute magnitude in the $NUV$ of TDEs measured by {\it Swift} follow-up of optically selected TDEs, in comparison to the mean UV properties at the optical peak of SNe Ia measured by {\it Swift} \citep{Milne2013}, and by the {\it GALEX} satellite for Type II-P SNe \citep{Gezari2015} and AGN \citep{Gezari2013}.  

It is striking how easily the UV-optical color alone can be used to filter our SNe.  With a wide-field, UV space telescope, observing concurrently with ground-based optical surveys, one could promptly distinguish TDEs from from SNe, and ``clear the fog" of contaminants, without the labor of follow-up spectroscopy, and the potential incompleteness introduced by filtering-down the full transient alert stream.  

\begin{figure}[t]
\includegraphics[width=3in,trim=1cm 0 4cm 14cm, clip]{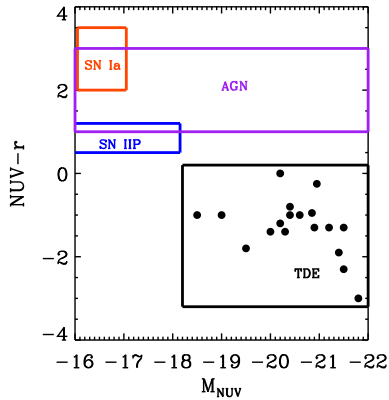}
\caption{\footnotesize Classification power of UV imaging to disentangle optical transients associated with AGN and SNe from bonafide TDEs, from their extremely blue UV-optical colors and high UV luminosities at peak.}
\label{fig:uvtde}
\end{figure}

There is still a possibility that searching for TDEs with the UV/optical properties of the optically-selected TDE candidates may miss an entire population of TDEs that are {\it not} UV-bright!  While X-ray follow-up has been possible for the promptly discovered optical TDEs, there has been much less comprehensive multiwavelength follow-up for the soft X-ray selected TDEs.   It may be that the component producing the UV/optical component, whether from stream-stream collisions, reprocessing, or a combination of both, is not present in all TDE systems.  A soft X-ray selected TDE sample, with contemporaneous optical monitoring, could address the following important questions:  \\
\indent $\bullet$ are optical and X-ray selected TDEs tracing the same population?  \\
\indent $\bullet$ what are the parameters that favor optical vs. X-ray emission?  \\
\indent $\bullet$ how does the optical and X-ray evolution of the TDEs constrain emission models?  \\
\indent $\bullet$ what are the timescales for the formation of an accretion disk in a TDE as a function of its black hole mass?  \\
\indent $\bullet$ what is the true TDE rate?  \\

With new X-ray survey capabilities, such as the extended Roengen Survey with an Imaging Telescope Array (eROSITA) on board the Russian-German Spektrum-Roentgen-Gamma (SRG) mission, currently conducting an all-sky X-ray (0.5-2 keV) survey, there should be soon a much larger sample (potentially thousands) of X-ray selected TDEs to shed light on these questions \citep{Khabibullin2014}.

An exciting X-ray signal that is predicted, but has yet to be detected, for stars on deep encounters at the moment of pericenter passage, is a "shock break out" when the shock produced from the tidal compression of the star propagates through and heats the deformed star \citep{Carter1983, Brassart2008}, resulting in a pulse of X-ray radiation, with a temperature of $k_{\rm B}T \sim \frac{GM_{\star} m_p}{R_{\star}} \sim 1 m_\star r_\star^{-1}$ keV, a duration of order fo the crossing time, $\delta t \sim R_{\star}/v_p \sim 10~m_\star^{-1/6} r_\star^{3/2} M_6^{-1/3}$ s, where $v_{p} \sim c(R_g/R_T)^{1/2}$ is the orbital velocity at periastron, and a luminosity of $L_{\rm X} \lsim 10^{42}~m_\star^{19/12} r_\star^{-5/4} M_6^{1/6}$ ergs s$^{-1}$ \citep{Kobayashi2004, Yalinewich2019}.
The detection of shock breakout would provide a clean time stamp for the time of disruption.  Detecting this brief X-ray burst is challenging, however, and requires a combination of wide area and high sensitivity in the X-ray band, potentially achievable with future X-ray telescopes being developed with lobster-eye X-ray optics \citep{Hudec2017}.

\subsection{Intermediate Mass Black Holes}
One of the most exciting prospects of using TDEs as probes of MBH demographics, is using the detection of the disruption of a white dwarf (WD) as a smoking-gun signature of an intermediate-mass black hole (IMBH), the elusive missing link between MBHs and stellar-mass black holes from the remnants of the evolution of massive stars (see recent review on IMBHs by \citep{Greene2019}).  The reason for this, is that given the high density of a WD, and the fact that the maximum mass for which a star is disrupted outside the event horizon scales as, $M_{\rm max} \propto R_\star^{3/2}/M_\star^{1/2} \propto \rho_\star^{1/2}$, WDs can only be disrupted by lower-mass black holes.   Assuming typical values of a C/O white dwarf, of $M_{\rm WD} \sim 0.6 M_\odot$ and $R_{\rm WD} \sim 0.014 R_\odot$, one gets $M_{\rm max} \sim 2 \times 10^{5} M_\odot$.  Furthermore, given the distinct radii and compositions of C/O WDs from He WDs, formed from the truncated evolution of red giants in a close binary system, one could use the observed TDE spectra to distinguish between them \citep{Law-Smith2017}, and get a more accurate measure of $M_{\rm BH}$ from the timescales of the system. 

 Indeed, there is at least one candidate for a WD disruption by an IMBH from the detection of [O~III]$\lambda 5007$ and [N~II]$\lambda 6583$ narrow-line emission, and no detection of hydrogen Balmer emission lines, from a globular cluster in the Fornax elliptical galaxy NGC 1399, which is also coincident with an ultra-luminous X-ray source \citep{Irwin2010}.  The strong oxygen line, and lack of any hydrogen Balmer-line emission, was interpreted as emission from photoionized hydrogen-free gas from a tidally disrupted star around an IMBH powering the X-ray emission via accretion.   Although the strong [N~III] emission is not predicted from photoionization models for the spectrum from a disrupated CO white dwarf \citep{Clausen2011}.  Unfortunately, there is no optical spectral information about the other IMBH candidate TDE detected by {\it Chandra} from a dwarf galaxy \citep{Maksym2013}, which would definitively identify it as a WD disruption and thus require an IMBH.
 
In general, a WD TDE will be a challenge to detect, since the characteristic timescales for a WD disruption very short, $t_{\rm fb} \sim 2~{\rm min}~(0.014/R_\odot)^{3/2} M_5^{1/2} (0.6/M_\odot)^{-1} $, and the peak luminosities lower, due to the lower Eddington luminosity of the IMBH involved.   
However, another potential manifestation of a WD disruption is the detonation of thermonuclear runaway from the tidal compression of the WD in a deep encounter with the IMBH, resulting in a supernova-like transient powered by the radioactive decay of iron-group elements from explosive nuclear burning \citep{Brassart2008, Rosswog2009, Anninos2018}.  This would result in a Type Ia supernova-like thermonuclear transient, albeit likely underluminous in comparison, due to the much different geometry and lower ejecta mass involved in the explosion \citep{MacLeod2016}.  

\subsection{TDEs as Multi-Messenger Sources}

\subsubsection{Very High Energy Neutrinos}
The first multi-messenger observation of a TDE is the IceCube detection of a very high energy (VHE) neutrino in the direction of a radio and X-ray detected optical TDE discovered by ZTF, AT2019dsg/ZTF19aapreis \citep{Stein2020}, establishing TDEs as a potential site for PeV neutrino production.  The detected $\sim 0.2$ PeV neutrino must have been produced by protons accelerated to high energies ($>$ 4 PeV) that then collided with a photon target (p$\gamma$ production) from the thermal TDE continuum, or a proton target (pp production), in the form of the unbound stellar debris or the radio-emitting $v \sim 0.1 c$ outflow.  No gamma-ray emission was detected by the Fermi-LAT telescope.  However, a coincident TeV gamma-ray signal, potentially detectable with gamma-ray Cherenkov telescopes, would confirm a hadronic origin, since the pp production produces a power-law neutrino spectrum, which would be accompanied by a lower-energy gamma rays.  Future multi-messenger observations of TDEs in PeV neutrinos, TeV neutrinos, and gamma rays, can distinguish between these neutrino production mechanisms, and be used to model the contribution of TDEs to the diffuse cosmic neutrino flux.  

\subsubsection{Gravitational Waves}

With the detection of the neutron star binary merger GW170817 in gravitational waves and in light across the electromagnetic spectrum, we officially entered a new era of multi-messenger astronomy \citep{Abbott2017}.  From this one source, one learned how combining information from gravitational waves and light can reveal profound insights into the physical properties of a transient like never before.  Since the the gravitational wave (GW) strain for a binary system scales as,

\begin{equation}
h \sim \frac{GM_\star R_g}{c^2 DR_p} \sim 2 \times 10^{-22} \beta \left ( \frac{D}{\rm 10~Mpc} \right)^{-1} r_\star^{-1} m_\star^{4/3} M_6^{2/3},
\end{equation}

\noindent with a frequency of 

\begin{equation}
f \sim \left ( \frac{GM_{\rm BH}}{R_p^3} \right )^{1/2} \sim 6 \times 10^{-4} \beta^{3/2} m_\star^{1/2} r_\star^{-3/2} {\rm Hz},
\end{equation}

\noindent the best chance of detecting a TDE in gravitational waves is from the tidal disruption of a compact star, like a white dwarf, on a deep encounter with a black hole, with a low frequency GW detector like the {\it Laser Interferometer Space Antenna} ({\it LISA}), but out to a distance of only 10 kpc \citep{Kobayashi2004, Rosswog2009, Anninos2018}.  This again favors the tidal disruption around an IMBH and is therefore an even more exciting as probe for IMBHs speculated to exist in dwarf galaxy nuclei and/or globular clusters \citep{Greene2019}.  With both a GW and EM signal, one could use the chirp mass measured from the GW signal, as well as the exact timing of the pericenter passage of the star, together with the light generated from the debris streams, to model the accretion flow in detail, and determine the true timescale for circularization and accretion \citep{Eracleous2019}.  
However, even without producing a detectable GW signal themselves, TDEs can provide a useful constraint on the rate of detecting GWs from extreme-mass ratio inspirals (EMRIs) of compact objects into massive black holes, which will be an important signal for {\it LISA}.  

\section{Conclusions}

In this review, I have highlighted the exciting progress in the discovery of TDEs from wide-field X-ray, UV, and optical surveys, including the build up of a statistically significant sample of TDEs with well-sampled light curves and detailed multi-wavelength characterization of their emission from the radio to the hard X-rays.  The panchromatic emission detected from TDEs probe a wide range of scales, from the innermost regions of a newly forming accreting debris disk, to the first collisions of debris streams as they fallback on the the black hole after disruption, to the expansion of the unbound debris, wind, or jet into the circumnuclear medium, and the heating of circumnuclear dust.  While TDEs are now well established as a class of nuclear transients with common, distinguishable properties, there are still important unknowns about what powers their luminous emission.  However, they certainly have realized their potential as probes of accretion onto $6 < \log (M_{\rm BH}/M_\odot) < 8$ black holes, and may be the only observable probes of accretion onto intermediate-mass ($\log (M_{\rm BH}/M_\odot) < 5$) black holes in the future.  

The most important observed trends can be summarized as follows.  

\noindent $\bullet$ The power-law decline rate of TDE light curves follows the expectation for emission that follows the fallback rate, $t_{\rm fb} \propto M_{\rm BH}^{1/2}$, but the rise time does not, and appears to be more closely related to the radiative diffusion timescale.   \\
$\bullet$ The characteristic radius of the UV/optical thermal emission in TDEs is consistent with the spatial scale of the intersection radius of the bound debris streams near apocenter, while the characteristic radius of the soft X-ray component is consistent with $R_{\rm isco}$ or even smaller.\\
$\bullet$ TDEs can be grouped into three spectral classes, TDE-H, TDE-H+He, and TDE-He, with the TDE-H+He class showing Bowen fluorescence line emission, and a preference for a hotter and more compact line-emitting region.  The TDE-He class is the rarest, and may be associated with the tidal disruption of evolved or stripped stars with helium-rich composition.\\
$\bullet$ The spectral line profiles of TDEs in the optical and UV are very broad, and with detailed photoionization modeling, can be a good tracer of the kinematics and geometry of the stellar debris disk and its associated wind.\\ 
$\bullet$ The UV to soft X-ray ratio in TDEs is highly variable, and appears to approach a ratio close to unity at late times.\  However, the lack of an decrease in observed line-of-sight absorption with increasing soft X-ray flux disfavors the "unveiling" of the soft X-ray emission through an expanding reprocessing layer.\\
$\bullet$ Infrared echoes and transient coronal-line emission are valuable indirect probes of TDEs from their influence on their gas and dust in the circumnuclear environment.\\
$\bullet$ There is strong evidence for outflows and/or jets powered by TDEs from the UV, radio, and X-rays.  Mapping their properties as a function of TDE parameters can provide insight on jet formation and AGN feedback.\\
$\bullet$ Observational constraints on the rate of TDEs is improving, and appear to be consistent with the expectations of stellar dynamical models.\\
$\bullet$ TDEs have a strong preference for galaxies in the green valley with centrally concentrated stellar density profiles, which may be the underlying cause for their frequent detection in "E+A" galaxy hosts.\\
$\bullet$ The future is bright for multi-wavelength and multi-messenger observations of TDEs, and their use as probes of central massive black hole demographics, accretion physics, and jet formation over cosmic time.\\

\section*{DISCLOSURE STATEMENT}
The author is not aware of any affiliations, memberships, funding, or financial holdings that
might be perceived as affecting the objectivity of this review. 

\section*{ACKNOWLEDGMENTS}
I want to thank my close collaborators in the observational hunt for TDEs from the ZTF "black hole" Science Working Group, especially Sjoert van Velzen, S. Brad Cenko, Matthew Graham, and Shri Kulkarni, and my terrific team of current and former graduate students, who have spent many nights observing, and days scanning for, TDEs: Tiara Hung, Sara Frederick, Charlotte Ward, and Erica Hammerstein.  I acknowledge support for some of this work from the National Science Foundation CAREER grant 1454816, and from several NASA {\it Neil Gehrels Swift} and {\it XMM-Newton} grants.  I also want to thank the Aspen Center for Physics, supported by National Science Foundation grant PHY-1607611, for hosting our 2018 Winter Workshop on ``Using Tidal Disruption Events to Study Super-Massive Black Holes", which was one of the first conferences dedicated to TDEs.  Finally, I want to thank my "village", Tanikwa, Gloria, Lexi, Natalie, my husband Chase, and The Sherwood Forest Boys and Girls Camp, without whose wonderful care of my adorable children, Cutter, Steele, Brooke and Ryder, I would not have been able to focus on finishing my review during a global pandemic.  
%

\bibliographystyle{ar-style2}
\bibliography{main}

\end{document}